\documentclass[a4paper,11pt]{article}                                                   

\usepackage[english]{babel}
\usepackage{a4wide}   

\usepackage{lipsum}
\usepackage{amsmath,amssymb,amsfonts}
\usepackage{graphicx}
\usepackage{epstopdf}
\usepackage{algorithmic}
\usepackage{dsfont}
\usepackage{mathrsfs}
\newcommand{\Tr}{{\rm Tr}}

\newtheorem{theorem}{Theorem}[section]
\newtheorem{lemma}[theorem]{Lemma}
\newtheorem{proposition}[theorem]{Proposition} 
\newtheorem{corollary}[theorem]{Corollary}
\newtheorem{remark}[theorem]{Remark}
\newtheorem{definition}[theorem]{Definition}

\title{\LARGE \bf
Exploring the Robustness of stabilizing controls for stochastic quantum evolutions
}
\date{\vspace{-5ex}}

\author{Weichao Liang\thanks{Laboratoire Math\'ematiques et Informatique pour la Complexit\'e et les Syst\`emes, CentraleSup\'elec, Universit\'e Paris Saclay
  (\it{weichao.liang@centralesupelec.fr}).}
\and Kentaro Ohki \thanks{Department of Applied Mathematics and Physics, Graduate School of Informatics, Kyoto University
  (\it{ohki@bode.amp.i.kyoto-u.ac.jp}).}
\and Francesco Ticozzi \thanks{Department of Information Engineering, University of Padova
  (\it{ticozzi@dei.unipd.it}).}}

\begin{document}

\maketitle

\newcommand{\FT}{\color{blue}}
\newcommand{\R}{\mathbb{R}}
\newcommand{\C}{\mathbb{C}}
\newcommand{\Hi}{\mathcal{H}}
\newcommand{\proof}{\textit{Proof.}}
\newcommand{\qed}{\hfill$\square$}

\begin{abstract}
 In this work we analyze and bound the effect of modeling errors on the stabilization of pure states or subspaces for quantum stochastic evolutions. Different approaches are used for open-loop and feedback control protocols. For both, we highlight the key role of dynamical invariance of the target: if the perturbation preserves invariance, it is possible to prove that it also preserves its attractivity, under some additional assumptions. In addition, we prove boundedness in mean of the solutions of perturbed systems under open-loop protocols. For the feedback strategies, in the general case without assumptions on invariance, we provide bounds on the perturbation effect in expectation and in probability, as well as specific bounds for non-demolition nominal systems. 
\end{abstract}

\section{Introduction}
The use of quantum control techniques has become increasingly common in a diverse array of experimental settings, including applications of quantum information, quantum computation, and quantum chemistry~\cite{altafini2012modeling}. In this work, focus on methods that aim to stabilize quantum systems towards a target pure state or subspace. Typical applications include entanglement generation and state preparation \cite{mirrahimi2007stabilizing,ticozzi2012hamiltonian}, quantum information encoding \cite{baggio} and protection \cite{feedbackQEC,gregoratti2004quantum}.  When the system is subjected to measurements, the general framework for studying these problems is based on Quantum Stochastic Master Equations (QSME)~\cite{belavkin1989nondemolition,bouten2007introduction,barchielli2009quantum}, where an open quantum system interacts with its environment and a probe system, with the latter continuously monitored. The conditioning on the measurement outcomes induces stochastic fluctuations of the best estimate of the state of the system. The control action is typically associated with a (possibly time-dependent) Hamiltonian perturbation or engineered dissipative dynamics \cite{mirrahimi2007stabilizing,ticozzi2012stabilization,grigoletto2021switching} 

A wide array of methods for control design have been proposed, but the robustness of these strategies remains a critical and largely unexplored issue for practical applications.  For example, most control design methods rely on the perfect knowledge of the model, and on the invariance of the target for the uncontrolled nominal dynamics.
However, in practice, one might have to take into account modeling errors, imperfect measurements, additional noise and couplings to the environment. In this work, we aim to characterize the cases in which the control strategy robustly attains the desired task with respect to the perturbations, and to provide bounds for the error when this is not the case.

The robustness of control laws has been studied mainly with respect to specific errors or models. These include uncertain model operators~\cite{petersen2012robust1,petersen2012robust2} and control input errors~\cite{ticozzi2004robust} based on robust control techniques. In \cite{van2007filtering,amini2014stability} and related papers, robustness of the filter with respect to initialization errors has been explored. Moreover, recent works~\cite{liang2021robust,liang2022model} proposed an explicit feedback controller that attains stabilization when the initial state and certain model parameters are unknown.  However, these robust control schemes assume perturbations with particular structures or parametrization.  

As realistic modeling errors may change or destabilize equilibria completely, we pursue a deeper analysis of the impact of more general, undesired modeling errors.  
We study the robustness of QSME stability with respect to three types of errors: (i) Hamiltonian perturbations, (ii) modeling errors in the measurement operators, (iii) additional un-modeled dissipative dynamics. The results are derived under minimal technical assumptions and bounds on the perturbation norms, avoiding particular parametrization when possible. 
We first focus on open-loop design methods, and then explore feedback strategies. The two scenarios call for different techniques, with the closed-loop case presenting more challenges as the dynamics becomes nonlinear. We provide here a brief outlook of the main results, which are:
\begin{description}
\item {\bf Proposition \ref{prop:genericgas}} shows that if the nominal system is Globally Exponentially Stable (GES), and the perturbations preserve invarance of the target for all values of their norms, a.s. GES is preserved (at least) for almost all values of the perturbations norms. The proof relies on linear-algebraic techniques and its details are presented in the appendix.
\item {\bf Proposition \ref{Prop:ISS}} bounds the error (in expectation and in probability) induced by perturbations that do not preserve invariance, leveraging stochastic analysis results;
\item {\bf Theorem \ref{Thm:GES Feedback}} shows that Hamiltonian and dissipative perturbations preserve the a.s. GES induced by a  feedback control strategies for {\em non-demolition} QSME if they preserve invariance and the measurement operators are accurately known;
\item{\bf Proposition  \ref{Prop:general_perturbation}} bounds the effect of general perturbation on the stability in probability of Globally Asymptotically Stable (GAS) dynamics. The result is actually a direct consequence of instrumental, general results derived for classical stochastic differential equations.
\end{description}

Some preliminary version of these results have been presented in the conference paper \cite{liang2023cdc}, where however we only considered the effect of errors of type (iii), and the feedback strategies have been studied only in the case of a particular spin system.

The paper is organized as follows: In Section~\ref{sec:SystemDescription}, we introduce the systems of interest, the stability notions and the problems we address. In Section~\ref{Sec:Open}, we analyze the effects of undesired Markovian couplings on the stability of quantum systems under open-loop protocols. 
In Section~\ref{Sec:Feedback}, we investigate the stability of perturbed quantum systems under feedback protocols. 
In Section~\ref{Sec:Conclusion}, we summarize our findings and discuss the implications of our work for future research.

\section{QSMEs and their stability}
\label{sec:SystemDescription}
\subsection{Notations and Models}

Let us briefly recall some notations that will be used throughout the paper.
We denote $\mathcal{B}(\mathcal{H})$ the set of all linear operators on a finite-dimensional Hilbert space $\mathcal{H}$, and define $\mathcal{B}_{*}(\mathcal{H}):=\{X\in\mathcal{B}(\mathcal{H})|X=X^*\}$, $\mathcal{B}_{\geq 0}(\mathcal{H}):=\{X\in\mathcal{B}(\mathcal{H})|X\geq 0\}$ and $\mathcal{B}_{>0}(\mathcal{H}):=\{X\in\mathcal{B}(\mathcal{H})|X> 0\}$. 
We use $\mathbf{I}$ to denote the identity operator on $\mathcal{H}$, and $\mathds{1}$ for indicator functions. 
We denote the adjoint $A\in\mathcal{B}(\mathcal{H})$ by $A^*$. As usual, the imaginary unit is denoted by $i$. For any finite positive integer $n$, we define $[n]:=\{1,\dots,n\}$.

The commutator and anticommutator of two operators $A,B\in\mathcal{B}(\mathcal{H})$ is denoted by $[A,B]:=AB-BA$ and $\{A,B\}=AB+BA$ respectively.
We denote $\bar{\boldsymbol\lambda}(A)$ and $\underline{\boldsymbol\lambda}(A)$ the maximum and minimum eigenvalue of Hermitian $A$, respectively. 
The function $\mathrm{Tr}(A)$ corresponds to the trace of $A\in\mathcal{B}(\mathcal{H})$. The Hilbert-Schmidt norm of $A\in\mathcal{B}(\mathcal{H})$ is denoted by $\|A\|:=\mathrm{Tr}(AA^*)^{1/2}$.  

We denote by $\mathrm{int}\{\mathcal{S}\}$ the interior of a subset of a topological space $\mathcal{S}$ and by $\partial \mathcal{S}$ its boundary. For $x\in\mathbb{C}$, $\mathbf{Re}\{x\}$ is the real part of $x$.

We consider quantum systems described on a finite-dimensional Hilbert space $\mathcal{H}$. The state of the system is associated to a density matrix on $\mathcal{H}$,  
$$
\mathcal{S}(\mathcal{H}):=\{\rho\in\mathcal{B}(\mathcal{H})|\,\rho=\rho^*\geq 0,\mathrm{Tr}(\rho)=1\}.
$$
Following~\cite{barchielli2009quantum}, we present the model for continuous-time quantum trajectories $\rho_t$.

Let $(\Omega, (\mathcal{F}_t)_t,\mathcal{F}_{\infty},\mathbb{Q})$ be a filtered probability space. We fix $n$ scalar $\mathbb{Q}$-Wiener processes $Y_k$ with $k\in [n]$, representing the outputs of a quantum system undergoing $k$ continuous time measurement of the homodyne/heterodyne type. Denote by $\mathcal{F}^Y_t:=\sigma(Y(s), 0\leq s \leq t)$ the filtration generated by the observation process up to time $t$.
Let $\rho_t$ denote the state of the system conditioned on measurement record, which obeys the Quantum Stochastic Master Equations (QSME) or Belavkin filtering equation:
\begin{align}
 d\rho_{t}=\mathcal{L}_{u}(\rho_{t})dt+\sum^n_{k=1}\sqrt{\eta_k}\mathcal{G}_{L_k}(\rho_{t})\big(dY_k(t)-\sqrt{\eta_k}\Tr((L_k^*+L_k)\rho_t)dt\big), \label{Eq:SME}
\end{align}
with $\rho_0 \in \mathcal{S}(\mathcal{H})$, where $\eta_k\in(0,1]$ represents the measurement efficiency, $\mathcal{L}_{u}(\rho):=-i[H_0+u_t H_1,\rho]+\sum^n_{k=1}\mathcal{D}_{L_k}(\rho)$ is called Lindblad generator, $\mathcal{D}_{L_k}(\rho) := L_k \rho L_k ^* - \frac{1}{2}L_k^* L_k \rho - \frac{1}{2} \rho L_k^* L_k$, and $\mathcal{G}_{L_k}(\rho):=L_k\rho+\rho L^*_k-\Tr((L_k+L_k^*)\rho)\rho$. 
Here $H_0,H_1\in\mathcal{B}_{*}(\mathcal{H})$ are the operators describing the free and control Hamiltonian of the system, and $L_k\in\mathcal{B}(\mathcal{H})$ is associated to the coupling with the probe field. If the operators described above correspond to the actual ones and no further couplings with the environment are relevant for the dynamics, the above equation represents the actual evolution of the systems. In the following, we will show that 
$Y_k(t)-\int^t_0\sqrt{\eta_k}\Tr((L_k^*+L_k)\rho_s)ds$ 
are also Wiener processes under a different probability measure.

While we consider to be able to access the same measurement records $Y_k$, we next assume that the actual operators and dynamics are different from the nominal values we have available. We suppose the dynamics is perturbed by Markovian couplings to additional external systems: the corresponding effect can be described by a sum of finite Lindblad generators $\gamma\sum^m_{k=1}\mathcal{D}_{C_k}(\rho)$ where $C_k\in\mathcal{B}(\mathcal{H})$ are the associated noise operators. 
Moreover, we suppose that our knowledge of the measurement is uncertain, in the sense that the actual noise operator induced by the interaction between the system and the probe is $\bar{L}_{k}=L_k+\beta\tilde{L}_{k}$. 
Lastly, we consider an actual effective Hamiltonian by $\bar{H}_0=H_0+\alpha \tilde{H}_0$. 

Then, the total dynamical perturbation can be described by the super-operator $F_{\alpha,\beta,\gamma}(\rho):=-i[\alpha\tilde{H}_0,\rho]+\sum^{n}_{k=1}\big(\mathcal{D}_{\bar{L}_k}(\rho)-\mathcal{D}_{L_k}(\rho)\big)+\gamma\sum^m_{k=1}\mathcal{D}_{C_k}(\rho)$. For the purpose of simplicity of the analysis, we assume that $\|\tilde{H}_0\|\leq 1$, $\|\tilde{L}_k\|\leq 1 $ for all $k$, and $\sum_k\|C_k\|\leq 1$, and the perturbation intensity is modulated through three parameters $\alpha,\beta\in\mathbb{R}$ and $\gamma\geq 0$. 
\begin{proposition}\label{Prop:Perturbed SME}
    The stochastic processes 
    $$
    \overline{W}_k(t) = Y_k(t)-\int^t_0\sqrt{\eta_k}\Tr((\bar{L}_k^*+\bar{L}_k)\sigma_s)ds, \quad k\in [n],
    $$ 
    are independent Wiener processes with respect to a probability measure $\overline{\mathbb{P}}^{\rho}$ that depend on the initial state $\rho,$ with the conditioned state now obeying the equation 
    \begin{align}
 d\sigma_{t}=\big(\mathcal{L}_{u}(\sigma_{t})+F_{\alpha,\beta,\gamma}(\sigma_{t})\big)dt+\sum^n_{k=1}\sqrt{\eta_k}\mathcal{G}_{\bar{L}_k}(\sigma_{t})d\overline{W}_k(t), \quad \sigma_0 = \rho \in \mathcal{S}(\mathcal{H}). \label{Eq:SME-P}
\end{align} 
   For any $\rho\in \mathcal{S}(\mathcal{H})$, with respect to $\overline{\mathbb{P}}^{\rho}$, Equation~\ref{Eq:SME-P} has a unique strong solution in $\mathcal{S}(\mathcal{H})$.
\end{proposition}
\proof In order to model the imperfect detection with the efficiency $\eta_k\in(0,1]$, we consider the corrupting noise by assuming that the probability space admits a $n$-dimensional process $(B_k(t))_{k\in[n]}$ which is independent of $(Y_k(t))_{k\in[n]}$, and set $\bar{Y}_k(t)=\sqrt{\eta_k}Y_k(t)+\sqrt{1-\eta_k}B_k(t)$. To describe the perturbed system, we consider the external perturbation as $m$ unobserved channels. These channels are associated with a $m$-dimensional measurement process $\hat{Y}(t)$, which is assumed to be admitted in the probability space.
Based on the actual noise operators $(\bar{L}_k)_{k\in[n]}$ induced by measurements and $\sqrt{\gamma}C_k$ induced by the external perturbation, and the actual effective Hamiltonian $\bar{H}_0$, we have following \cite{barchielli2009quantum} that the dynamical propagator is associated to the following matrix-valued stochastic differential equation
\begin{align*}
    d\mathsf{S}_t = -\big(i(\bar{H}_0+u_t H_1)+\tfrac{1}{2}&\textstyle\sum^n_{k=1}\bar{L}^*_k\bar{L}_k+\tfrac{\gamma}{2}\textstyle\sum^m_{j=1}C^*_kC_k\big)\mathsf{S}_tdt\nonumber\\
    &+\textstyle\sum^n_{k=1}\bar{L}_k\mathsf{S}_td\bar{Y}_k(t)+\textstyle\sum^m_{j=1}\sqrt{\gamma}C_j\mathsf{S}_td\hat{Y}_j(t),\quad \mathsf{S}_0=\mathbf{I}.
\end{align*}
For any initial condition $\rho\in\mathcal{S}(\mathcal{H})$, define the unnormalized state for the system as $\bar{\varsigma}_t:=\mathbb{E}_{\mathbb{Q}}(\mathsf{S}_t \rho \mathsf{S}^*_t|\mathcal{F}^Y_t)$, where $\mathbb{E}_{\mathbb{Q}}$ denotes the expectation corresponding to $\mathbb{Q}$.

Then, take $\mathbb{E}_{\mathbb{Q}}(\cdot|\mathcal{F}^Y_t)$ on two sides of the above equation. Due to linearity, independence between the processes, and the elementary results on the conditional expectation and Fubini's theorem~\cite[Lemma 5.4]{xiong2008introduction},
it is possible then to show that:
\begin{equation*}
    d\bar{\varsigma}_t = -i[\bar{H}_0+u_t H_1,\bar{\varsigma}_t]dt + \sum^n_{k=1}\mathcal{D}_{\bar{L}_k}(\bar{\varsigma}_t)dt+\gamma\sum^m_{j=1}\mathcal{D}_{C_j}(\bar{\varsigma}_t)dt + \sum^n_{k=1}\sqrt{\eta_k}(\bar{L}_k\bar{\varsigma}_t + \bar{\varsigma}_t \bar{L}_k^*)dY_k(t)
\end{equation*}
with $\bar{\varsigma}_0 = \rho$, and $\mathsf{Z}^{\rho}_t:=\Tr(\bar{\varsigma}_t)$ is the unique solution of the following positive real-valued stochastic differential equation
\begin{equation*}
    d\mathsf{Z}^{\rho}_t =  \sum^n_{k=1}\sqrt{\eta_k}\Tr(\bar{L}_k\bar{\varsigma}_t + \bar{\varsigma}_t \bar{L}_k^*)dY_k(t) = \mathsf{Z}^{\rho}_t\sum^n_{k=1}\sqrt{\eta_k}\Tr\big((\bar{L}_k + \bar{L}_k^*)\sigma_t\big)dY_k(t), \quad \mathsf{Z}^{\rho}_0 = 1,
\end{equation*}
where $\sigma_t:=\bar{\varsigma}_t/\mathsf{Z}^{\rho}_t$. Moreover, $\mathsf{Z}^{\rho}_t$ is a non-negative $\mathbb{Q}$-martingale~\cite[Theorem 3.4]{barchielli2009quantum}. 

For any $\rho\in\mathcal{S}(\mathcal{H})$, we define a probability $\overline{\mathbb{P}}^\rho_t$ on $(\Omega,\mathcal{F}_t)$: $d\overline{\mathbb{P}}^\rho_t=\mathsf{Z}^{\rho}_t d\mathbb{Q}|_{\mathcal{F}_t}$. Since $\mathsf{Z}^{\rho}_t$ is a non-negative $\mathbb{Q}$-martingale, the family $(\overline{\mathbb{P}}^{\rho}_t)_t$ is consistent. This defines a unique probability $\overline{\mathbb{P}}^{\rho}$ on $(\Omega,\mathcal{F}_{\infty})$. We will denote by $\overline{\mathbb{E}}^{\rho}$ the expectation with respect to $\overline{\mathbb{P}}^{\rho}$. Then, by applying the Girsanov theorem, we have that 
\begin{equation}\label{Eq:Girsanov}
    \overline{W}_k(t) = Y_k(t)-\int^t_0\sqrt{\eta_k}\Tr\big((\bar{L}_k^*+\bar{L}_k)\sigma_s\big)ds, \quad k\in [n],
\end{equation}
    are independent Wiener processes with respect to $\overline{\mathbb{P}}^{\rho}$. 
By applying the It\^o's formula, we can obtain the QSME in terms of $\overline{W}_k(t)$ in the form \eqref{Eq:SME-P}. The rest of the statement follows from the derivation above.
\qed
 
Notice that \eqref{Eq:SME-P} is still of the form \eqref{Eq:SME}, up to the addition of the perturbations and a change of probability measure. The same construction and results of the proposition above apply to \eqref{Eq:SME} with the {\em nominal} values of the parameters, yet with respect to a different measure.
Consider $Y_k(t), Z^\rho_t$ and $\mathbb{Q}$ as above: for any $\rho\in\mathcal{S}(\mathcal{H})$, we define a probability $\mathbb{P}^\rho_t$ on $(\Omega,\mathcal{F}_t)$: $d\mathbb{P}^\rho_t=Z^{\rho}_t d\mathbb{Q}|_{\mathcal{F}_t}$. Since $Z^{\rho}_t$ is a nonnegative $\mathbb{Q}$-martingale, the family $(\mathbb{P}^{\rho}_t)_t$ is consistent, that is $\mathbb{P}^{\rho}_t(E)=\mathbb{P}^{\rho}_s(E)$ for $t\geq s$ and $E\in\mathcal{F}_s$. This defines a unique probability $\mathbb{P}^{\rho}$ on $(\Omega,\mathcal{F}_{\infty})$ by Kolmogorov's extension theorem. We will denote by $\mathbb{E}^{\rho}$ the expectation with respect to $\mathbb{P}^{\rho}$. Then, by applying the Girsanov theorem as above, we have that the stochastic processes 
    $$
    W_k(t) = Y_k(t)-\int^t_0\sqrt{\eta_k}\Tr\big((L_k^*+L_k)\rho_s\big)ds, \quad k\in [n],
    $$ 
    are independent Wiener processes with respect to $\mathbb{P}^{\rho}$. 
We refer to~\cite{barchielli2009quantum} for a complete reference on the physical interpretation of the above-mentioned Wiener processes.

We can thus obtain exactly the QSME \eqref{Eq:SME} using the same construction as in the proposition, this time in terms of $W_k(t)$, which describes the time evolution of the conditioned quantum state associated to the {\em nominal} parameters of the QSME.

\subsection{Stability notions}
Let $\mathcal{H}_S\subset \mathcal{H}$ be the target subspace, i.e., the subspace towards which we would like to converge. Denote by $\Pi_{0}\notin\{0,\mathbf{I}\}$ the orthogonal projection on $\mathcal{H}_S\subset \mathcal{H}$. Define the set of density matrices
$
\mathcal{I}(\mathcal{H}_S):=\{\rho\in\mathcal{S}(\mathcal{H})| \mathrm{Tr}(\Pi_{0}\rho)=1\},
$
namely those whose support is contained in $\mathcal{H}_S$.
\begin{definition}
For a system of the form ~\eqref{Eq:SME}, the subspace $\mathcal{H}_S$ is called invariant almost surely if $\rho_0\in \mathcal{I}(\mathcal{H}_S)$, $\rho_t\in \mathcal{I}(\mathcal{H}_S)$ for all $t>0$ almost surely.
\end{definition}

Based on the notions of stochastic stability defined in~\cite{khasminskii2011stochastic,mao2007stochastic} and the definition used in~\cite{ticozzi2008quantum,benoist2017exponential}, we collect the key definitions of interest for our work.

\begin{definition}
Let $\mathcal{H}_S\subset\mathcal{H}$ be an invariant subspace for the system~\eqref{Eq:SME}, and denote by $\Pi_{0}$ the orthogonal projection on $\mathcal{H}_S$ and $\mathbf{d}_{0}(\rho):=\|\rho-\Pi_{0}\rho\Pi_{0}\|$. Then $\mathcal{H}_S$ is said to be
\begin{enumerate}

\item
\emph{stable in mean}, if for every $\varepsilon \in (0,1)$, there exists $\delta = \delta(\varepsilon)>0$ such that,
\begin{equation*}
\mathbb{E} \big( \mathbf{d}_{0}(\rho_t) \big) \leq \varepsilon,
\end{equation*}
whenever $\mathbf{d}_{0}(\rho_0)<\delta$. 

\item \emph{globally asymptotically stable (GAS) in mean}, if it is stable in mean and,
\begin{equation*}
\mathbb{E} \left( \lim_{t\rightarrow\infty}\mathbf{d}_{0}(\rho_t)\right) = 0, \quad \forall \rho_0\in\mathcal{S}(\mathcal{H}).\end{equation*}

\item
\emph{globally exponentially stable (GES) in mean}, if there exist a pair of positive constants $\lambda$ and $c$ such that
\begin{equation*}
\mathbb{E}\big(\mathbf{d}_{0}(\rho_t)\big)\leq c \,\mathbf{d}_{0}(\rho_0) e^{-\lambda t},\quad \forall \rho_0\in\mathcal{S}(\mathcal{H}).
\end{equation*}

\item
\emph{stable in probability}, if for every $\varepsilon \in (0,1)$ and for every $r>0$, there exists $\delta = \delta(\varepsilon,r)>0$ such that,
\begin{equation*}
\mathbb{P} \big( \mathbf{d}_{0}(\rho_t)< r \text{ for } t \geq 0 \big) \geq 1-\varepsilon,
\end{equation*}
whenever $\mathbf{d}_{0}(\rho_0)<\delta$. 

\item 
\emph{globally asymptotically stable (GAS) almost surely}, if it is stable in probability and,
\begin{equation*}
\mathbb{P} \left( \lim_{t\rightarrow\infty}\mathbf{d}_{0}(\rho_t)=0 \right) = 1, \quad \forall \rho_0\in\mathcal{S}(\mathcal{H}).
\end{equation*}

\item
\emph{globally exponentially stable (GES) almost surely}, if
\begin{equation*}
\limsup_{t \rightarrow \infty} \frac{1}{t} \log \big( \mathbf{d}_{0}(\rho_t)\big) < 0,\quad \forall \rho_0\in\mathcal{S}(\mathcal{H}), \quad a.s.
\end{equation*}
The left-hand side of the above inequality is called the \emph{sample Lyapunov exponent}.
\end{enumerate}
\end{definition}

In the paper \cite{ticozzi2012stabilization,benoist2017exponential} the notions of stability in mean, in probability and a.s. have been shown to be equivalent for time-invariant systems in the form \eqref{Eq:SME}, so the same is true for the case of time-invariant open-loop control.

Based on the nature of the control input $u_t$, there are two types of control protocols for the stabilization of the system~\eqref{Eq:SME}: 1) if $u_t$ is a real bounded deterministic process depending on $t$, this protocol is called open-loop stabilization~\cite{benoist2017exponential,ticozzi2012stabilization}.  Notice that for time-invariant, linear systems, stability in mean is always exponential. This is the case when we consider open-loop, constant controls. 2) if $u_t$ is a real bounded stochastic process depending on the measurement output up to $t$, this protocol is called feedback stabilization~\cite{mirrahimi2007stabilizing,liang2021robust}. 

In this paper, we suppose the target subspace $\mathcal{H}_S\subset \mathcal{H}$ is GES almost surely with respect to the nominal system~\eqref{Eq:SME}, and we analyze the effect of the perturbation on the stability of the nominal system under open-loop and feedback protocols respectively.

\subsection{Conditions for robust invariance}
Let $\mathcal{H}=\mathcal{H}_S\oplus\mathcal{H}_R$ and $X\in\mathcal{B}(\mathcal{H})$, the matrix representation of $X$ in a basis obtained joining basis of $\mathcal{H}_S$ and $\mathcal{H}_R$ can be written as 
\begin{equation*}
X=\left[\begin{matrix}
X_S & X_P\\
X_Q & X_R
\end{matrix}\right],
\end{equation*}
where $X_S,X_R,X_P$ and $X_Q$ are matrices representing operators from $\mathcal{H}_S$ to $\mathcal{H}_S$, from $\mathcal{H}_R$ to $\mathcal{H}_R$, from $\mathcal{H}_R$ to $\mathcal{H}_S$, from $\mathcal{H}_S$ to $\mathcal{H}_R$, respectively. We denote by $\Pi_R$ the orthogonal projection on $\mathcal{H}_R\subset\mathcal{H}$. In~\cite{ticozzi2008quantum}, Ticozzi and Viola showed that the invariance of the subspace enforces a given structure for the associated semi-group generator by the following lemma. 
\begin{lemma}\label{Lem: Invariant}
    For the nominal system~\ref{Eq:SME}, the subspace $\mathcal{H}_S$ is invariant in mean and almost surely if and only if
    $$
    \forall k\in[n],\quad L_{k,Q}=0 \text{ and } i(H_{0,P}+u_t H_{1,P})=\frac{1}{2}\sum^n_{k=1}L^*_{k,S}L_{k,P}.
    $$
\end{lemma}
The above conditions, written for the perturbed dynamics, imply that if $\mathcal{H}_S$ is invariant for the nominal dynamics \eqref{Eq:SME}, it also remains invariant in mean and almost surely with respect to the perturbed system~\eqref{Eq:SME-P}, if and only if the following condition holds. 
\begin{itemize}
    \item[\textbf{A}:] $\forall k\in[n]$, $\forall j\in[m]$, $\tilde{L}_{k,Q}=C_{j,Q}=0$ and 
    $$
    2\alpha i\tilde{H}_{0,P}=\beta\sum^n_{k=1}\big(L^*_{k,S}\tilde{L}_{k,P}+\tilde{L}^*_{k,S}L_{k,P}+\beta \tilde{L}^*_{k,S}\tilde{L}_{k,P}\big) +\gamma\sum^m_{j=1} C^{*}_{j,S}C_{k,P}.
    $$
\end{itemize}
{However, from a practical perspective, it is unlikely that a parametric perturbation could satisfy \textbf{A} only for a set of particular sets of values of $\alpha,\beta,\gamma$. For this reason, we introduce a sufficient, stronger condition which implies the previous and guarantees invariance for any choice of the parameters:
\begin{itemize}
    \item[\textbf{AR}:] $\forall k\in[n]$, $\forall j\in[m]$,
    \begin{align*}
        & \tilde{L}_{k,Q} = C_{j,Q} = 0, \\
        & \tilde{H}_{0,P} = 0, \\
        & \textstyle\sum_{k=1}^n \big(L^*_{k,S} \tilde{L}_{k,P} + \tilde{L}^*_{k,S} L_{k,P}\big) = 0, \\
        & \textstyle\sum_{k=1}^n \big(\tilde{L}^*_{k,S} \tilde{L}_{k,P}\big) = 0, \\
        & \textstyle\sum_{j=1}^m C^*_{j,S} C_{j,P} = 0.
    \end{align*}
\end{itemize}

\section{Robustness of open-loop stabilization}
\label{Sec:Open}

In this section, we suppose that $u_t\equiv u$ where $u$ is a real constant and $\mathcal{H}_S$ is GES almost surely with respect to the nominal system~\eqref{Eq:SME}. 

We define the following maps,
\begin{align*}
    \mathcal{L}_{R}(\rho_R)&:=-i[H_R,\rho_R]+\textstyle\sum^{n}_{k=1}\mathfrak{D}_{L_k}(\rho_R),\\
    \overline{\mathcal{L}}_{R}(\rho_R)&: = -i[\bar{H}_R,\rho_R]+\textstyle\sum^{n}_{k=1}\mathfrak{D}_{\bar{L}_k}(\rho_R)+\gamma\textstyle\sum^{m}_{j=1}\mathfrak{D}_{C_k}(\rho_R),
\end{align*}
where $H_R:=H_{0,R}+u H_{1,R}$, $\bar{H}_R:=H_R+\alpha\tilde{H}_{0,R}$, $\mathfrak{D}_{A}(\rho_R):=A_{R}\rho_R A^*_{R}-\frac{1}{2}\{A^*_{P}A_{P}+A^*_{R}A_{R},\rho_R\}$, and $\rho_R\in\mathcal{B}_{\geq 0}(\mathcal{H}_R)$. For $R$-block of any Lindblad generator, we denote $\mathcal{L}^*_{R}$ and $\overline{\mathcal{L}}^*_{R}$ the adjoint of $\mathcal{L}_{R}$ and $\overline{\mathcal{L}}_{R}$ respectively with respect to the Hilbert-Schmidt inner product on $\mathcal{B}(\mathcal{H}_R)$. 
In the following, we denote by $\bar{\lambda}_{\alpha,\beta,\gamma}$ the spectral abscissa of $\overline{\mathcal{L}}_{R}$, i.e., $\bar{\lambda}_{\alpha,\beta,\gamma}:=\min\{-\mathbf{Re}(x)|\,x\in\mathrm{sp}(\overline{\mathcal{L}}_{R})\}$ where the parameters $(\alpha,\beta,\gamma)$ are those appearing in $\overline{\mathcal{L}}_{R}$. Note that the spectral abscissa of $\mathcal{L}_{R}$, $\lambda=\lim_{\alpha,\beta,\gamma\rightarrow 0}\bar{\lambda}_{\alpha,\beta,\gamma}$ since $\mathcal{L}_{R}=\lim_{\alpha,\beta,\gamma\rightarrow 0}\overline{\mathcal{L}}_{R}$.

\subsection{Perturbations that preserve invariance}
Firstly, we consider the case where $\mathcal{H}_S$ is still invariant for the perturbed system~\eqref{Eq:SME-P}. Then, the results in~\cite[Section 2]{benoist2017exponential} imply directly the following.
\begin{proposition}\label{Prop:OpenGES}
    Suppose that 
    \emph{\textbf{A}} is satisfied.
    For any $\varepsilon>0$, there exists $K_{R}\in\mathcal{B}_{>0}(\mathcal{H}_R)$ such that 
    $
    \overline{\mathcal{L}}_{R}^*(K_{R})\leq -(\lambda-\varepsilon)K_{R}.
    $
    Moreover, $\mathcal{H}_S$ is GES almost surely for the perturbed system~\eqref{Eq:SME-P} if and only if $\bar{\lambda}_{\alpha,\beta,\gamma}>0$.
\end{proposition}
The above proposition requires full information on the perturbation to ensure the GES of target subspace. In the following, we consider the case where $\mathcal{H}_S$ is GES almost surely for the nominal system~\eqref{Eq:SME}, that is $\lambda>0$ by Proposition~\ref{Prop:OpenGES}. 

\begin{corollary}
 Suppose $\lambda>0$. Then, there exist $K_R\in\mathcal{B}_{>0}(\mathcal{H}_R)$ and $c>0$ such that $\mathcal{L}^*_{R}(K_R)\leq -c K_R$. Moreover, if \emph{\textbf{AR}} holds and $c>R_{\alpha,\beta,\gamma}\|K_R\|/\underline{\boldsymbol\lambda}(K_R)$ with 
$R_{\alpha,\beta,\gamma}:=2\big(\alpha+\beta\sum^n_{k=1}(2\|L_{k,R}\|+\|L_{k,P}\|)+n\beta^2+\gamma \big)$,
 then $\mathcal{H}_S$ is GES almost surely for the perturbed system~\eqref{Eq:SME-P}.
\end{corollary}
\textit{Proof.}
The first part can be obtained directly by applying Proposition~\ref{Prop:OpenGES}. For any $\alpha\geq 0$, we have
\begin{align*}
    \overline{\mathcal{L}}^*_{R}(K_R)= \mathcal{L}^*_{R}(K_R)+\mathsf{F}_{\alpha,\beta,\gamma}(K_R)\leq -cK_R+\mathsf{F}_{\alpha,\beta,\gamma}(K_R),
\end{align*}
where
\begin{align*}
    \mathsf{F}_{\alpha,\beta,\gamma}(K_R):=\alpha [K_R,\tilde{H}_{0,R}]&+\beta\sum^n_{k=1} \mathsf{D}_{\bar{L}_k}(K_R)+\beta^2\sum^n_{k=1} \mathfrak{D}^*_{\tilde{L}_k}(K_R)+\gamma\sum^m_{k=1} \mathfrak{D}^*_{C_k}(K_R),
\end{align*}
with $\mathsf{D}_{\bar{L}_k}(K_R):=\tilde{L}_R^*K_RL_R+L_RK_R\tilde{L}_R-\frac{1}{2}\{\tilde{L}^*_PL_p+L^*_P\tilde{L}_P+\tilde{L}^*_RL_R+L^*_R\tilde{L}_R,K_R\}$.
Since $K_R>0$, we have 
\begin{align*}
    \mathsf{F}_{\alpha,\beta,\gamma}(K_R)\leq \frac{\bar{\boldsymbol\lambda}\big(\mathsf{F}_{\alpha,\beta,\gamma}(K_R)\big)}{\underline{\boldsymbol\lambda}(K_R)}K_R\leq \frac{\|\mathsf{F}_{\alpha,\beta,\gamma}(K_R)\|}{\underline{\boldsymbol\lambda}(K_R)}K_R.
\end{align*}
Due to the relation $\|AB\|\leq\|A\|\|B\|$ and the assumption $\|\tilde{H}_0\|\leq 1$, $\|\tilde{L}_k\|\leq 1$ for all $k\in[n]$ and $\sum_k\|C_k\|\leq 1$, we have
$
    \|\mathsf{F}_{\alpha,\beta,\gamma}(K_R)\|\leq R_{\alpha,\beta,\gamma}\|K_R\|.
$
Therefore, we deduce,
\begin{align*}
    \overline{\mathcal{L}}^*_{R}(K_R)\leq -\Big( c-\frac{R_{\alpha,\beta,\gamma}\|K_R\|}{\underline{\boldsymbol\lambda}(K_R)}\Big)K_R.
\end{align*}
The results can be concluded by applying similar arguments as in~\cite[Section 2]{benoist2017exponential}.
\hfill$\square$

Suppose now the conditions \textbf{AR} are satisfied. Based on \textit{Dissipation - Induced Decomposition} technique \cite{ticozzi2012hamiltonian}, we can establish that in fact $\mathcal{H}_S$ remains GES almost surely for the perturbed system~\eqref{Eq:SME-P} with at most the exception of a set of zero measure (in the Lebesgue sense). The necessary tools and the proof of the key instrumental result are presented in the appendix, in order to avoid a rather lengthy detour. We can summarize the relevant conclusion in the following.
\begin{proposition}\label{prop:genericgas}
     Suppose that $\lambda>0$ and \emph{\textbf{AR}} is satisfied. Then, $\mathcal{H}_S$ remains GES in mean and almost surely for the perturbed system~\eqref{Eq:SME-P} for almost all values of $\alpha,\beta,\gamma.$
\end{proposition}

\textit{Proof.}
For constant $u$, the results of \cite{benoist2017exponential}  ensure that proving GES in mean is sufficient to prove GES both in mean and almost surely for the perturbed system, and that GAS in mean is equivalent to GES in mean. The result then follows from a direct application of Theorem \ref{thm:genericgas}, which is proved in the appendix, with $x=(\alpha,\beta,\gamma).$ 
\hfill$\square$

\subsection{Effect of general perturbations}
Next, we consider the general case, the perturbation is supposed to be unknown. In this case, GES should not be expected since $\mathcal{H}_S$ may not be invariant for the perturbed system. Instead, we study the boundedness of $\overline{\mathbb{E}}^{\sigma}[\mathbf{d}_{0}(\sigma(t))]$ for any $\sigma\in\mathcal{S}(\mathcal{H})$.

\begin{proposition} \label{Prop:ISS}
    Suppose that $\lambda>0$. Then, there exist $K_R\in\mathcal{B}_{>0}(\mathcal{H}_R)$ and $c>0$ such that $\mathcal{L}^*_{R}(K_R)< -c K_R$. Denote the {\em extension of $K_R$ to} $\mathcal{B}(\mathcal{H})$ by
$
K=\left[\begin{smallmatrix}
0 & 0\\
0 & K_R
\end{smallmatrix}\right].
$
Then, for any $\sigma\in\mathcal{S}(\mathcal{H})$, for the perturbed system~\eqref{Eq:SME-P} with $\sigma_0=\sigma,\, t\geq 0$,
    \begin{equation}     
    \overline{\mathbb{E}}^{\sigma}[\mathbf{d}_{0}(\sigma(t))]\leq \sqrt{\tfrac{3}{\underline{\boldsymbol\lambda}(K_R)}\Big(\Tr(K\sigma)e^{-ct}+D_{\alpha,\beta,\gamma}\big(1- e^{-ct}\big)/c \Big)},    \label{Eq:boundedness}
    \end{equation}
    and for all $\delta\geq 1$, 
     \begin{equation}\label{Eq:ISS}
        \overline{\mathbb{P}}^{\sigma}\Big[\mathbf{d}_{0}(\sigma(t))< \delta \sqrt{\tfrac{3}{\underline{\boldsymbol\lambda}(K_R)}\big(\Tr(K\sigma)e^{-ct}+D_{\alpha,\beta,\gamma}(1- e^{-ct})/c \big) }\Big]\geq 1-1/\delta,
    \end{equation}
    with $\mathfrak{L}:=\sum^n_{k=1}\|L_k\|$, and $D_{\alpha,\beta,\gamma}:=2\|K_R\|( |\alpha|+n|\beta|^2+2|\beta|\mathfrak{L}+\gamma).$
\end{proposition}
\textit{Proof.}
Note that $\Tr(K\sigma)=\Tr(K_R\sigma_R)$. Due to the linearity, 
we have 
\begin{align*}
   \mathscr{L}\Tr(K\sigma)&=\Tr\big(K\mathcal{L}_{u}(\sigma)+KF_{\alpha,\beta,\gamma}(\sigma)\big)\\
   & = \Tr\big(\mathcal{L}^*_{R}(K_R)\sigma_R\big)+\Tr\big(KF_{\alpha,\beta,\gamma}(\sigma)\big)\\
   &\leq -c\Tr(K_R\sigma_R)+\big|\Tr\big(KF_{\alpha,\beta,\gamma}(\sigma)\big)\big|\\
   &\leq -c\Tr(K\sigma)+D_{\alpha,\beta,\gamma},
\end{align*}
where $\mathscr{L}$ is related to the perturbed system~\eqref{Eq:SME-P} and $D_{\alpha,\beta,\gamma}:=2\|K_R\|( |\alpha|+n|\beta|^2+2|\beta|\mathfrak{L}+\gamma)$ with $\mathfrak{L}:=\sum^n_{k=1}\|L_k\|$, we used the Cauchy–Schwarz inequality, the relation $\|AB\|\leq\|A\|\|B\|$ and $\|\sigma_R\|\leq \|\sigma\|\leq 1$ and the assumption $\|\tilde{H}_0\|\leq 1$, $\|\tilde{L}_k\|\leq 1$ for all $k\in[n]$ and $\sum_k\|C_k\|\leq 1$. 

By applying the It\^o's formula and then taking the expectation, for any initial state $\sigma_0=\sigma\in\mathcal{S}(\mathcal{H})$, we get 
$$
\frac{d}{dt} e^{ct}\overline{\mathbb{E}}^{\sigma}(\Tr(K\sigma_t))\leq  D_{\alpha,\beta,\gamma} e^{ct},
$$
which implies 
$$
\overline{\mathbb{E}}^{\sigma}(\Tr(K\sigma_t))\leq \Tr(K\sigma)e^{-ct}+D_{\alpha,\beta,\gamma}\big(1- e^{-ct}\big)/c.
$$
Next, by a straightforward computation, we find that $\mathbf{d}_{0}(\sigma)^2=\|\sigma_R\|^2+2\|\sigma_P\|^2$ where $\|\sigma_R\|^2\leq \Tr(\Pi_R\sigma)$. Since $\sigma\geq 0$ and $\Tr(\sigma)=1$, we have $\|\sigma_P\|^2\leq \Tr(\Pi_S\sigma)\Tr(\Pi_R\sigma)\leq \Tr(\Pi_R\sigma)$. Thus, it follows $\mathbf{d}_{0}(\sigma)^2\leq 3 \Tr(\Pi_R\sigma)\leq \frac{3}{\underline{\boldsymbol\lambda}(K_R)} \Tr(K\sigma)$.
This leads to the equation~\eqref{Eq:boundedness}. Moreover, for any $\delta\geq 1$ and $\sigma_0=\sigma\in \mathcal{S}(\mathcal{H})$, we have
\begin{align*}
    \overline{\mathbb{E}}^{\sigma}[\mathbf{d}_{0}(\sigma_t)]\geq \overline{\mathbb{E}}^{\sigma}[\mathbf{d}_{0}(\sigma_t)\mathds{1}_{\{\mathbf{d}_{0}(\sigma_t)\geq \delta f_{\sigma}(t)\}}]\geq  \delta f_{\rho}(t) \overline{\mathbb{P}}^{\sigma}\big(\mathbf{d}_{0}(\sigma_t)\geq \delta f_{\sigma}(t)\big),
\end{align*}
where $f_{\sigma}(t):=\sqrt{\tfrac{3}{\underline{\boldsymbol\lambda}(K_R)}\big(\Tr(K\sigma)e^{-ct}+D_{\alpha,\beta,\gamma}(1- e^{-ct})/c \big)}>0$. The latter, together with the inequality~\eqref{Eq:boundedness}, yields equation~\eqref{Eq:ISS}.
\hfill$\square$
\begin{remark}
The property~\eqref{Eq:boundedness} ensures that the solution of the perturbed system~\eqref{Eq:SME-P} is bounded in mean, moreover, when the perturbation vanishes, i.e., $\alpha=\beta=\gamma=0$, the system~\eqref{Eq:SME-P} is GES in mean, then by following the similar arguments as in~\cite{benoist2017exponential} we can show it is also GES almost surely. 
The property~\eqref{Eq:ISS} is referred to as the \textit{stochastic noise-to-state stability}~\cite{liu2008notion,ito2019complete}. 
In our case stability is also exponential.
\end{remark}

\section{Robustness of feedback stabilization under quantum non-demolition measurements}
\label{Sec:Feedback}
In this section, we explore the robustness of the state feedback stabilization strategy provided in~\cite{liang2020robustness} for the system~\eqref{Eq:SME-P}.
We shall assume that the dynamics has a generalized non-demolition form: consider a decomposition of the whole Hilbert space: $\mathcal{H}=\mathcal{H}_S\oplus\mathcal{H}^1_{R}\oplus \dots \oplus \mathcal{H}^d_{R}$. Denote by $\Pi_0,\Pi_1,\dots,\Pi_d$ the orthogonal projections on $\mathcal{H}_S,\mathcal{H}^1_{R}, \dots,\mathcal{H}^d_{R}$ respectively. Thus, the projections resolve the identity, i.e., $\sum^d_{j=0}\Pi_j=\mathbf{I}$.

We then assume  $H_0=\sum^d_{j=0}h_j \Pi_j$ with $h_j\in \mathbb{R}$ and $L_k=\sum^d_{j=0}l_{k,j}\Pi_j$ with $l_{k,j}\in\mathbb{C}$ for all $k\in[n]$. That is, they are simultaneously block-diagonalizable with respect to the decomposition above.
Moreover, we impose the following assumptions on the noise operators $L_k$ induced by the measurements:
\begin{itemize}
    \item[\textbf{A1.1}:] For each $k\in[n]$, $\mathbf{Re}\{l_{k,i}\}$ for all $i\in\{0,\dots,d\}$ cannot be identical.
    \item[\textbf{A1.2}:] For all $j\in[d]$, $\sum^n_{k=1}|\mathbf{Re}\{l_{k,j}\}-\mathbf{Re}\{l_{k,0}\}|>0$. 
\end{itemize}
Furthermore, we limit ourselves to the case where the noise operator induced by measurements $L_k$ is proportionally perturbed, i.e., $\bar{L}_k = \sqrt{\theta_k}L_k$ with $\theta_k>0$ for all $k\in [n]$. It can be written in the form: $\bar{L}_k = L_k+\beta_k \tilde{L}_k$ with $\tilde{L}_k=L_k/\|L_k\|$ and $\beta_k=(\sqrt{\theta_k}-1)\|L_k\|$. This is a technical assumption that is crucial in deriving the main result: it indicates that a good knowledge of the measurement operators is key to robust stability. 
The dynamics of perturbed system are given by the following stochastic master equation:
\begin{align}
 d\sigma_{t} = \big(-i[H_0+u_t H_1,\sigma_t]+&\textstyle\sum^n_{k=1}\theta_k\mathcal{D}_{L_k}(\sigma_t)+F_{\alpha,\gamma}(\sigma_{t})\big) dt\nonumber\\
 &+\textstyle\sum^n_{k=1}\sqrt{\eta_k \theta_k}\mathcal{G}_{L_k}(\sigma_{t})d\overline{W}_k(t), \quad \sigma_0 =\sigma \in \mathcal{S}(\mathcal{H}). \label{Eq:SME-Pf}
\end{align} 
where $F_{\alpha,\gamma}(\sigma)=F_{\alpha,0,\gamma}(\sigma)$.

From a practical point of view, the initial state of the system $\sigma$, the measurement efficiency $\eta_k$ and the perturbation cannot be precisely known, following the treatments in~\cite{liang2020robustness} and Proposition~\ref{Prop:Perturbed SME}, we construct an estimated state $\hat{\sigma}_t$ using
\begin{align*}
d\hat{\sigma}_t=&\Big(-i[H_0+u_t H_1,\hat{\sigma}_t]+\textstyle\sum^n_{k=1}\hat{\theta}_k\mathcal{D}_{L_k}(\hat{\sigma}_t)\Big) dt\\
&+\textstyle\sum^n_{k=1}\sqrt{\hat{\eta}_k\hat{\theta}_k}\mathcal{G}_k(\hat{\sigma}_t)\Big(dY_k(t)-\sqrt{\hat{\eta}_k\hat{\theta}_k}\Tr\big((L_k+L_k^*)\hat{\sigma}_t\big)dt\Big), \quad \hat{\sigma}_0=\hat{\sigma}\in\mathcal{S}(\mathcal{H}),
\end{align*}
where $\hat{\eta}_k\in(0,1]$ and $\hat{\theta}_k>0$ are used to denote the best available estimates for the parameters $\eta_k$ and $\theta_k$. Due to the relation~\eqref{Eq:Girsanov}, we have
\begin{align}
d\hat{\sigma}_t=\Big(-i[H_0&+u_t H_1,\hat{\sigma}_t]+\textstyle\sum^n_{k=1}\hat{\theta}_k\mathcal{D}_{L_k}(\hat{\sigma}_t)\Big) dt\nonumber\\
&+\textstyle\sum^n_{k=1}\sqrt{\hat{\eta}_k\hat{\theta}_k}\mathcal{G}_{L_k}(\hat{\sigma}_t)\Big(d\overline{W}_k(t)+\mathcal{T}_k\big(\sigma_t,\hat{\sigma}_t\big)dt\big), \label{Eq:SME-F}
\end{align}
where $\mathcal{T}_k(\sigma,\hat{\sigma}):=\sqrt{\eta_k \theta_k}\mathrm{Tr}((L_k+L^*_k)\sigma)-\sqrt{\hat{\eta}_k \hat{\theta}_k} \mathrm{Tr}((L_k+L^*_k)\hat{\sigma})$. The control input is a function of the estimator, i.e., $u_t=u(\hat{\sigma}_t)$, which is applied in the system~\eqref{Eq:SME-Pf} and the estimator~\eqref{Eq:SME-F}.

\subsection{Perturbations that preserve invariance}
It is direct to verify that $\mathcal{H}_S$ is invariant for the nominal system. Under the specific perturbation in this section, the sufficient conditions \textbf{AR} ensuring the invariance of $\mathcal{H}_S$ reduce to the following
\begin{itemize}    
\item[\textbf{AR'}:]  $\forall j\in[m]$, $C_{j,Q}=0$, $\tilde{H}_{0,P}=0,$ and $\sum^m_{j=1} C^{*}_{j,S}C_{k,P}=0.$
\end{itemize}
Denote $B_{\varepsilon}(\mathcal{H}_S):=\{\rho\in\mathcal{S}(\mathcal{H})|\mathbf{d}_0(\rho)<\varepsilon\}$ with $\varepsilon>0$. Now, we introduce the following assumptions on the feedback controller and control Hamiltonian to ensure $\mathcal{H}_S$ is the only invariant subspace of the estimator~\eqref{Eq:SME-F}.
\begin{itemize}
    \item[\textbf{H1}:] $u\in \mathcal{C}^1(\mathcal{S}(\mathcal{H}),\mathbb{R})$, $u(\hat{\sigma})=0$ for all $\hat{\sigma}\in B_{\kappa}(\mathcal{H}_S)$ for some $\kappa>0$ and $u(\hat{\sigma})\neq 0$ for all $\hat{\sigma}\in\mathcal{I}(\mathcal{H}^j_R)$  for all $j\in[d]$,
    \item[\textbf{H2}:] $\hat{H}^{(j)}_{1,P}$ is full rank, for all $j\in[d]$,
\end{itemize}
where the $\hat{H}^{(j)}_{1,P}$ is a block of $H_1$ representing the operator from $\mathcal{H}^{j}_{R}$ to $\mathcal{H}^{j-1}_{R}$ where we set $\mathcal{H}^{0}_{R}:=\mathcal{H}_S$, see Appendix~\ref{Sec:DID} for more details. 
The assumption $u\in \mathcal{C}^1(\mathcal{S}(\mathcal{H}),\mathbb{R})$ is used to ensure the existence and uniqueness of the solution of the coupled system~\eqref{Eq:SME-Pf}--\eqref{Eq:SME-F}, that together with the Feller continuity and the strong Markov property can be proved by the same arguments of \cite{mirrahimi2007stabilizing}. The almost sure invariance of $\mathcal{S}(\mathcal{H})\times \mathcal{S}(\mathcal{H})$ for \eqref{Eq:SME-Pf}--\eqref{Eq:SME-F} can be shown by the similar arguments as in Proposition~\ref{Prop:Perturbed SME}. \textbf{H2} is the sufficient condition to guarantee $[H_1,\hat{\sigma}]\neq 0$ for all $\hat{\sigma}\in\mathcal{I}(\mathcal{H}^j_R)$ for all $j\in [d]$. It is worth noting that the assumption \textbf{H2} is relatively strong compared to the conventional assumption in the literature. We can be mitigate the stringency of \textbf{H2} by carefully designing the control Hamiltonian, such considerations are beyond the scope of this paper and are addressed for our future work. Then, together with \textbf{H1}, we can ensure that $\mathcal{H}_S$ is the only invariant subspace among the subspaces $\mathcal{H}_S,\mathcal{H}^1_{R},\dots,\mathcal{H}^d_{R}$.

Define 
\begin{equation*}
    \bar{\mathfrak{c}}_k:=\mathbf{Re}\{l_{k,0}\}-\min_{j\in[d]}\mathbf{Re}\{l_{k,j}\}, \quad \underline{\mathfrak{c}}_k:=\mathbf{Re}\{l_{k,0}\}-\max_{j\in[d]}\mathbf{Re}\{l_{k,j}\},\quad \forall k\in[n].
\end{equation*}
We impose the following condition on the noise operators $L_k$ induced by the measurements:
\begin{itemize}
    \item[\textbf{A1.3}:] For each $k\in[n]$, $\bar{\mathfrak{c}}_k\leq 0$ or $\underline{\mathfrak{c}}_k\geq 0$.
\end{itemize}
This assumption is crucial for showing the recurrence relative to any neighborhood of $\mathcal{I}(\mathcal{H}_S)$ (Proposition~\ref{Prop:Recurrence}) and providing an estimation of the Lyapunov exponent.
In addition, while assessing the recurrence property, we may meet the case where the coupled system~\eqref{Eq:SME-Pf}--\eqref{Eq:SME-F} includes invariant subsets other than the target subset $\mathcal{I}(\mathcal{H}_S)\times \mathcal{I}(\mathcal{H}_S)$ (see Appendix~\ref{Sec:Recurrence} for the detailed exploration.). To ensure the instability of these non-desired invariant subsets (Lemma~\ref{Lem:Instability}), we introduce the parameter $\chi_k:= \sqrt{\eta_k \theta_k/\hat{\eta}_k \hat{\theta}_k}$ for all $k\in [n]$, and propose the following condition:
\begin{itemize}
        \item[\textbf{C1}:] For all $k\in [n]$, $\chi_k>1/2$, and $$\mathbf{Re}\{l_{k,0}\}<(2\chi_k-1)(\mathbf{Re}\{l_{k,0}\}-\bar{\mathfrak{c}}_k)  \text{ if } \bar{\mathfrak{c}}_k\leq0,$$  while  $$\mathbf{Re}\{l_{k,0}\}>(2\chi_k-1)(\mathbf{Re}\{l_{k,0}\}-\underline{\mathfrak{c}}_k) \text{ if } \underline{\mathfrak{c}}_k\geq 0.$$ 
    \end{itemize}
Refer to Remark~\ref{Rem:instability} for a exploration of assumption \textbf{A1.3} and condition \textbf{C1}, which offers further insights behind these choices.  

For any $\xi\in\mathcal{H}$ and $l\in\mathbb{Z}^{+}$, define the following matrix
$$
\mathbf{M}_{l,\xi}:=[\xi, H_1 \xi,L^*_1H_1\xi,\dots, L^*_nH_1\xi,\dots,H_1^l\xi,L^*_1H_1^l\xi,\dots, L^*_n H_1^l\xi].
$$

To ensure the non-invariance of $\{\sigma\in\mathcal{S}(\mathcal{H})|\,\Tr(\Pi_0\sigma)=0\}$ (Proposition~\ref{Prop:Recurrence}), we make the following assumption:
\begin{itemize}
    \item[\textbf{A2}:] Let $\Xi_S$ be a set of basis vectors of $\mathcal{H}_S$, there exists $l\in\mathbb{Z}^{+}$ such that, for all $\xi\in\Xi_S$, $\mathrm{rank}(\mathbf{M}_{l,\xi})\geq \mathrm{dim}(\mathcal{H})-1$.
\end{itemize}
In the following, we state our main results on the almost sure exponential stabilization of the perturbed system~\eqref{Eq:SME-Pf} via state feedback when the target subspace $\mathcal{H}_S$ is invariant. 

Define $\underline{\mathfrak{l}}_k:=\min_{j\in[d]}\big\{R_{k,j}\big|R_{k,j}:=|\mathbf{Re}\{l_{k,j}\}-\mathbf{Re}\{l_{k,0}\}|>0 \big\}$, which is well-defined provided condition \textbf{A1.1} is satisfied, and define $\bar{\mathfrak{l}}_k:=\max_{j\in[d]}\{|\mathbf{Re}\{l_{k,j}\}-\mathbf{Re}\{l_{k,0}\}|\}$ for all $k\in[n]$. Then, introduce two coefficients related to the estimation of the Lyapunov exponent,
\begin{align*}
     &\mathsf{C}:=\frac{1}{2}\min_{k\in[n]}\big\{\hat{\eta}_k\hat{\theta}_k\underline{\mathfrak{l}}_k^2\min\{\chi_k^2,1\}\big\}-2\mathfrak{C},\quad \mathsf{K}:=\frac{1}{2d}\min_{k\in[n]}\big\{\hat{\eta}_k\hat{\theta}_k\underline{\mathfrak{l}}_k^2\min\{\chi_k^2,1\}\big\},
\end{align*}
where $\mathfrak{C}:=\sum^n_{k=1}\hat{\eta}_k\hat{\theta}_k\bar{\mathfrak{l}}_k|\mathbf{Re}\{l_{k,0}\}(\chi_k-1)|$. Moreover, we impose the following condition ensuring $\mathsf{C}>0$, which guarantees the local stability in probability of the target subspace,
\begin{itemize}
        \item[\textbf{C2}:] $\min_{k\in[n]}\big\{\hat{\eta}_k\hat{\theta}_k\underline{\mathfrak{l}}_k^2\min\{\chi_k^2,1\}\big\}>4\mathfrak{C}$.
\end{itemize}
Condition \textbf{C2} provides a quantitative measure of robustness for the parameters. In the case where $n=1$, \textbf{C2} simplifies to $\underline{\mathfrak{l}}^2\min\{\chi^2,1\}>4\bar{\mathfrak{l}}|\mathbf{Re}\{l_{0}\}||\chi-1|$, which can be further interpreted in the following two cases: when $\mathbf{Re}\{l_{0}\}=0$, for all $\chi>0$, \textbf{C2} is satisfied; when $\mathbf{Re}\{l_{0}\}\neq 0$, for all $\chi\in(\underline{\chi},\bar{\chi})$, condition \textbf{C2} is satisfied, where $\underline{\chi}=\frac{2}{\underline{\mathfrak{l}}^2}\Big(\sqrt{(\underline{\mathfrak{l}}^2+\bar{\mathfrak{l}}|\mathbf{Re}\{l_{0}\}|)\bar{\mathfrak{l}}|\mathbf{Re}\{l_{0}\}|}-\bar{\mathfrak{l}}|\mathbf{Re}\{l_{0}\}|\Big)\in(0,1)$ and $\bar{\chi}=1+\underline{\mathfrak{l}}^2/4\bar{\mathfrak{l}}|\mathbf{Re}\{l_{0}\}|>1$.
\begin{theorem}\label{Thm:GES Feedback}
    Suppose that $\eta_k<1$ for all $k\in[n]$ and the assumptions \emph{\textbf{AR'}}, \emph{\textbf{H1}}, \emph{\textbf{H2}}, \emph{\textbf{A1.1}}-\emph{\textbf{A1.3}}, \emph{\textbf{A2}}, and the conditions \emph{\textbf{C1}} and \emph{\textbf{C2}} are satisfied. 
    Then, for all initial state $\sigma\in\mathcal{S}(\mathcal{H})$, the target subspace $\mathcal{H}_S$ is $\overline{\mathbb{P}}^{\sigma}$-almost sure GES for the perturbed system~\eqref{Eq:SME-Pf}, for almost all values of $(\alpha,\gamma)$ containing $(0,0)$, with the sample Lyapunov exponent less than or equal to $-\mathsf{C}-\mathsf{K}$.
\end{theorem}
\proof
The proof proceeds in three steps:
\begin{enumerate}
    \item First, we show that the trajectories $(\sigma_t,\hat{\sigma}_t)$ of the coupled system~\eqref{Eq:SME-Pf}--\eqref{Eq:SME-F} is recurrent relative to any neighborhood of $\mathcal{I}(\mathcal{H}_S)\times \mathcal{I}(\mathcal{H}_S)$;
    \item Next, we show the target subspace $\mathcal{H}_S$ is local stable in probability;
    \item Finally, we show the almost sure GAS and provide and an estimation of the Lyapunov exponent.
\end{enumerate}

\textit{Step 1.} From Proposition~\ref{Prop:Recurrence}, the recurrence property is ensured, that is, for any initial state $(\sigma_0,\hat{\sigma}_0)\in\mathcal{S}(\mathcal{H})\times\mathrm{int}\{\mathcal{S}(\mathcal{H})\}$, the trajectories $(\sigma_t,\hat{\sigma}_t)$ enters any neighborhood of $\mathcal{I}(\mathcal{H}_S)\times \mathcal{I}(\mathcal{H}_S)$ in finite time almost surely for almost all values of $(\alpha, \gamma)$ containing $(0,0)$.

\textit{Step 2.}
Consider the candidate Lyapunov function
$$
V(\sigma,\hat{\sigma})=\sum^d_{j=1}\sqrt{\mathrm{Tr}(\sigma \Pi_j)+\mathrm{Tr}(\hat{\sigma} \Pi_j)}\geq 0,
$$
where equality holds if and only if $(\sigma,\hat{\sigma})\in \mathcal{I}(\mathcal{H}_S)\times \mathcal{I}(\mathcal{H}_S)$. Due to Lemma~\ref{Lemma:PosDef invariant}, if $\hat{\sigma}_0>0$, then $\hat{\sigma}_t>0$ for all $t\geq 0$, $\overline{\mathbb{P}}^{\rho}$-almost surely. Consequently, it follows that $\mathrm{Tr}(\hat{\sigma_t} \Pi_j)>0$ for all $j\in[d]$, which further implies that $V(\sigma_t,\hat{\sigma}_t)>0$ for all $t\geq 0$, $\overline{\mathbb{P}}^{\sigma}$-almost surely.

Define $\mathbf{P}_{k,j}(\sigma):=\mathbf{Re}\{l_{k,j}\}-\sum^d_{i=0}\mathbf{Re}\{l_{k,i}\}\Tr(\Pi_i \sigma)$.
For all $(\sigma,\hat{\sigma})\in B_{\kappa}(\mathcal{H}_S)\times B_{\kappa}(\mathcal{H}_S)$ where $\kappa>0$ is defined in \textbf{H1}, the infinitesimal generator related to~\eqref{Eq:SME-Pf}--\eqref{Eq:SME-F} is given by
\begin{align*}
    \mathscr{L}& V(\sigma,\hat{\sigma})=\sum^d_{j=1}\big(\mathfrak{F}_j(\sigma,\hat{\sigma})-\mathfrak{G}_j(\sigma,\hat{\sigma})\big),
\end{align*}
where
\begin{align*}
    \mathfrak{F}_j(\sigma,\hat{\sigma})&:=\frac{\mathrm{Tr}(\hat{\sigma} \Pi_j)\sum^n_{k=1}\sqrt{\hat{\eta}_k\hat{\theta}_k}\mathbf{P}_{k,j}(\hat{\sigma})\mathcal{T}_k(\sigma,\hat{\sigma})}{\sqrt{\mathrm{Tr}(\sigma \Pi_j)+\mathrm{Tr}(\hat{\sigma} \Pi_j)}},\\
    \mathfrak{G}_j(\sigma,\hat{\sigma})&:=\frac{\sum^n_{k=1}\hat{\eta}_k\hat{\theta}_k\big(\chi_k\mathbf{P}_{k,j}(\sigma) \mathrm{Tr}(\sigma \Pi_j)+\mathbf{P}_{k,j}(\hat{\sigma})\mathrm{Tr}(\hat{\sigma} \Pi_j)\big)^2}{2\big(\mathrm{Tr}(\sigma \Pi_j)+\mathrm{Tr}(\hat{\sigma} \Pi_j)\big)^{3/2}}.
\end{align*}
We deduce that $|\mathbf{P}_{k,j}(\hat{\sigma})|\leq \bar{\mathfrak{l}}_k+|\mathbf{P}_{k,0}(\hat{\sigma})|$ for all $\hat{\sigma}\in\mathcal{S}(\mathcal{H})$ and $k\in[n]$. Consequently, we obtain
\begin{align*}
    \mathfrak{F}_j(\sigma,\hat{\sigma})&\leq \frac{\big(\mathrm{Tr}(\sigma \Pi_j)+\mathrm{Tr}(\hat{\sigma} \Pi_j)\big)\sum^n_{k=1}\sqrt{\hat{\eta}_k\hat{\theta}_k}|\mathbf{P}_{k,j}(\hat{\sigma})\mathcal{T}_k(\sigma,\hat{\sigma})|}{\sqrt{\mathrm{Tr}(\sigma \Pi_j)+\mathrm{Tr}(\hat{\sigma} \Pi_j)}}\\
    &\leq \sqrt{\mathrm{Tr}(\sigma \Pi_j)+\mathrm{Tr}(\hat{\sigma} \Pi_j)}\sum^n_{k=1}\sqrt{\hat{\eta}_k\hat{\theta}_k}(\bar{\mathfrak{l}}_k+|\mathbf{P}_{k,0}(\hat{\sigma})|)|\mathcal{T}_k(\sigma,\hat{\sigma})|.
\end{align*}
For all $k\in[n]$, we have $\mathbf{P}_{k,j}(\sigma)=\mathbf{Re}\{l_{k,j}\}-\mathbf{Re}\{l_{k,0}\}+\mathbf{P}_{k,0}(\sigma)$. 
Under assumptions \textbf{A1.1} and \textbf{A1.2}, there exists at least one $\mathsf{k}\in[n]$ such that $\mathbf{Re}\{l_{\mathsf{k},j}\}\neq \mathbf{Re}\{l_{\mathsf{k},0}\}$ for all $j\in[d]$. Moreover, since $\lim_{\sigma\rightarrow\mathcal{I}(\mathcal{H}_S)}\mathbf{P}_{k,0}(\sigma)=0$ for all $k\in[n]$, for all $\sigma\in B_{\zeta}(\mathcal{H}_S)$ with $\zeta>0$ sufficiently small, we have 
\begin{equation*}
    \begin{cases}
       \mathbf{P}_{\mathsf{k},j}(\sigma) >0,& \text{if }\mathbf{Re}\{l_{\mathsf{k},j}\}>\mathbf{Re}\{l_{\mathsf{k},0}\};\\
       \mathbf{P}_{\mathsf{k},j}(\sigma) <0,& \text{if }\mathbf{Re}\{l_{\mathsf{k},j}\}<\mathbf{Re}\{l_{\mathsf{k},0}\}.
    \end{cases}
\end{equation*}
It follows that 
\begin{align*}
    \mathfrak{G}_j(\sigma,\hat{\sigma})&\geq \frac{\hat{\eta}_{\mathsf{k}}\hat{\theta}_{\mathsf{k}}\min\{\chi_{\mathsf{k}}^2\mathbf{P}_{\mathsf{k},j}(\sigma)^2,\mathbf{P}_{\mathsf{k},j}(\hat{\sigma})^2\} \big( \mathrm{Tr}(\sigma \Pi_j)+\mathrm{Tr}(\hat{\sigma} \Pi_j)\big)^2}{2\big(\mathrm{Tr}(\sigma \Pi_j)+\mathrm{Tr}(\hat{\sigma} \Pi_j)\big)^{3/2}}\\
    &=\sqrt{\mathrm{Tr}(\sigma \Pi_j)+\mathrm{Tr}(\hat{\sigma} \Pi_j)}\frac{\hat{\eta}_{\mathsf{k}}\hat{\theta}_{\mathsf{k}}}{2}\min\{\chi_{\mathsf{k}}^2\mathbf{P}_{\mathsf{k},j}(\sigma)^2,\mathbf{P}_{\mathsf{k},j}(\hat{\sigma})^2\}. 
\end{align*}
Additionally, we establish the following relation for all $k\in[n]$,
\begin{equation}\label{Eq:ineqP}
    \begin{cases}
     \mathbf{P}_{k,0}(\sigma)\leq \bar{\mathfrak{c}}_k (1-\Tr(\Pi_0\sigma))\leq 0,&   \text{if }\bar{\mathfrak{c}}_k\leq 0;\\
     \mathbf{P}_{k,0}(\sigma)\geq \underline{\mathfrak{c}}_k (1-\Tr(\Pi_0\sigma))\geq 0,& \text{if }\underline{\mathfrak{c}}_k\geq 0.
    \end{cases}
\end{equation}
Hence, under assumption \textbf{A1.3}, for all $\sigma\in B_{\zeta}(\mathcal{H}_S)$ with $\zeta>0$ sufficiently small, we have
$$
|\mathbf{P}_{\mathsf{k},j}(\sigma)|^2=|\mathbf{Re}\{l_{\mathsf{k},j}\}-\mathbf{Re}\{l_{\mathsf{k},0}\}+\mathbf{P}_{\mathsf{k},0}(\sigma)|^2\geq (\underline{\mathfrak{l}}_\mathsf{k}-|\mathbf{P}_{\mathsf{k},0}(\sigma)|)^2.
$$
This implies that 
\begin{align*}
    \mathfrak{G}_j(\sigma,\hat{\sigma})\geq \sqrt{\mathrm{Tr}(\sigma \Pi_j)+\mathrm{Tr}(\hat{\sigma} \Pi_j)}\frac{\hat{\eta}_{\mathsf{k}}\hat{\theta}_{\mathsf{k}}}{2}\min\{\chi_{\mathsf{k}}^2(\underline{\mathfrak{l}}_{\mathsf{k}}-|\mathbf{P}_{\mathsf{k},0}(\sigma)|)^2,(\underline{\mathfrak{l}}_{\mathsf{k}}-|\mathbf{P}_{\mathsf{k},0}(\hat{\sigma})|)^2\}.
\end{align*}
These inequalities imply that, for all $(\sigma,\hat{\sigma})\in B_{\zeta}(\mathcal{H}_S)\times B_{\zeta}(\mathcal{H}_S)$ with $\zeta>0$ sufficiently small,
\begin{align*}
    \mathscr{L} V(\sigma,\hat{\sigma}) \leq -\mathbf{C}(\sigma,\hat{\sigma})V(\sigma,\hat{\sigma}),
\end{align*}
where 
\begin{align*}
    \mathbf{C}(\sigma,\hat{\sigma}):=\frac{1}{2}\min_{k\in[n]}\big\{\hat{\eta}_k\hat{\theta}_k\min\{\chi_k^2(\underline{\mathfrak{l}}_k-|\mathbf{P}_{k,0}(\sigma)|)^2,(\underline{\mathfrak{l}}_k-|\mathbf{P}_{k,0}(\hat{\sigma})|)^2\}\big\}-\\
    \sum^n_{k=1}\sqrt{\hat{\eta}_k\hat{\theta}_k}(\bar{\mathfrak{l}}_k+|\mathbf{P}_{k,0}(\hat{\sigma})|)|\mathcal{T}_k(\sigma,\hat{\sigma})|.
\end{align*}

By a straightforward computation, we have
$$
    \lim_{(\sigma,\hat{\sigma})\rightarrow\mathcal{I}(\mathcal{H}_S)\times \mathcal{I}(\mathcal{H}_S)}\mathbf{C}(\sigma,\hat{\sigma})=\mathsf{C}>0,
$$
where the positivity of $\mathsf{C}$ is ensured by the condition \textbf{C2}.
Therefore, we have 
$$
\limsup_{(\sigma,\hat{\sigma})\rightarrow \mathcal{I}(\mathcal{H}_S)\times \mathcal{I}(\mathcal{H}_S)}\frac{\mathscr{L} V(\sigma,\hat{\sigma})}{V(\sigma,\hat{\sigma})}\leq \lim_{(\sigma,\hat{\sigma})\rightarrow \mathcal{I}(\mathcal{H}_S)\times \mathcal{I}(\mathcal{H}_S)}-\mathbf{C}(\sigma,\hat{\sigma}) = - \mathsf{C} < 0.
$$
 Due to the continuity of $\mathbf{C}(\sigma,\hat{\sigma})$, there exists a $\lambda\in (0,\kappa)$ such that 
$$
\mathscr{L} V(\sigma,\hat{\sigma}) \leq 0, \quad \forall (\sigma,\hat{\sigma})\in B_{\lambda}(\mathcal{H}_S)\times B_{\lambda}(\mathcal{H}_S),
$$
then by applying the similar arguments as in the proof of~\cite[Theorem 6.3]{liang2019exponential}, then local stability in probability is ensured.

\textit{Step 3.} Combining the results in \textit{Step 1} and \textit{Step 2}, by employing the similar arguments as in the proof of~\cite[Theorem 6.3]{liang2019exponential}, $\mathcal{H}_S\times \mathcal{H}_S$ is $\overline{\mathbb{P}}^{\rho}$-almost surely asymptotically stable with the initial condition $(\sigma_0,\hat{\sigma}_0)\in\mathcal{S}(\mathcal{H})\times\mathrm{int}\{\mathcal{S}(\mathcal{H})\}$. Moreover, we have 
\begin{align*}
    &\liminf_{(\sigma,\hat{\sigma})\rightarrow \mathcal{I}(\mathcal{H}_S)\times \mathcal{I}(\mathcal{H}_S)} \sum^n_{k=1}\Tr\Bigg( \frac{\partial V(\sigma,\hat{\sigma})}{\partial \sigma}\frac{\sqrt{\eta_k \theta_k}\mathcal{G}_{L_k}(\sigma)}{V(\sigma,\hat{\sigma})}+\frac{\partial V(\sigma,\hat{\sigma})}{\partial \hat{\sigma}}\frac{\sqrt{\hat{\eta_k} \hat{\theta_k}}\mathcal{G}_{L_k}(\hat{\sigma})}{V(\sigma,\hat{\sigma})} \Bigg)^2
    \\
    & \geq 2\mathsf{K}.
\end{align*}
By using arguments as in the proof of~\cite[Theorem 6.3]{liang2019exponential} again, we have
$$
\limsup_{t\rightarrow \infty} \frac{1}{t} \log  V(\sigma_t,\hat{\sigma}_t) \leq -\mathsf{C}-\mathsf{K}, \quad \overline{\mathbb{P}}^{\sigma}\text{-}a.s.
$$
Moreover, since $\mathbf{d}_0(\sigma)\leq \sqrt{3(1-\Tr(\Pi_0\sigma))}\leq \sqrt{3} V(\sigma,\hat{\sigma})$  which is established in the proof of Proposition~\ref{Prop:ISS}, it follows that
$$
\limsup_{t\rightarrow \infty}\frac{1}{t}\log  \mathbf{d}_0(\sigma_t) \leq -\mathsf{C}-\mathsf{K}, \quad \overline{\mathbb{P}}^{\sigma}\text{-}a.s.
$$
that completes the proof.
\hfill$\square$

\begin{remark}
   Theorem~\ref{Thm:GES Feedback} provides sufficient conditions for ensuring almost sure GES. However, we have not fully optimized the assumptions and the Lyapunov function to maximize the estimation of the Lyapunov exponent in this context. The assumptions \textbf{\emph{H1}} and the range of estimated parameters \textbf{\emph{C1}} and \textbf{\emph{C2}}, can be further relaxed and improved through more refined computations. For an example of such relaxation and improvement, refer to~\cite{liang2021robust,liang2022model}.
\end{remark}

As an example of application of the previous results, we consider the following feedback laws satisfying \textbf{H1}. Define a continuously differentiable function $f:[0,1]\to[0,1]$,
\begin{equation*}
f(x) := 
\begin{cases}
0,&\text{if }x\in[0,\varepsilon_1);\\
\frac12\sin\left(\frac{\pi(2x-\varepsilon_1-\varepsilon_2)}{2(\varepsilon_2-\varepsilon_1)}\right)+\frac12,&\text{if }x\in[\varepsilon_1,\varepsilon_2);\\
1,&\text{if }x\in(\varepsilon_2,1],
\end{cases}
\end{equation*}
where $0<\varepsilon_1<\varepsilon_2<1$. Define 
\begin{equation}
u(\hat{\sigma})=a \big(1-\mathrm{Tr}(\hat{\sigma} \Pi_0)\big)^{b}f(1-\mathrm{Tr}(\hat{\sigma} \Pi_0)),
\label{Eq:u}
\end{equation}
with $a>0$ and $b \geq1$, then \textbf{H1} holds true.

\subsection{Behavior of the system under general perturbations}
Let $(\rho_t,\hat{\rho}_t)$ be the solution of coupled system~\eqref{Eq:SME-Pf}--\eqref{Eq:SME-F} with $(\alpha,\gamma)=(0,0)$, and $(\sigma_t,\hat{\sigma}_t)$ be the solutions of the perturbed coupled system~\eqref{Eq:SME-Pf}--\eqref{Eq:SME-F} under the general perturbation, i.e., without assuming \textbf{AR'}, with $\rho_0=\sigma_0=\sigma\in\mathcal{S}(\mathcal{H})$ and $\hat{\rho}_0=\hat{\sigma}_0\in\mathrm{int}\{\mathcal{S}(\mathcal{H})\}$.

In the following proposition, we provide an estimation of $\overline{\mathbb{E}}^{\sigma}(\|\rho_t-\sigma_t\|)$ in finite time horizon.
It specifies the power rate of convergence, and the rate of getting to infinity of the lengths of the time interval. Both rates depend on the perturbation magnitude $\alpha\in\mathbb{R}$ and $\gamma\geq 0$.
\begin{proposition} \label{Prop:Stability}
    Suppose that the assumption \emph{\textbf{H1}} is satisfied. 
    Then, for any initial state $\sigma\in\mathcal{S}(\mathcal{H})$, there exist two constants $A,B>0$ such that, 
    $$\overline{\mathbb{E}}^{\sigma}(\|\rho_t-\sigma_t\|)\leq (|\alpha| + \gamma) A (e^{Bt}-1).$$
    Moreover, for any $\delta\in(0,1)$,
    $$
    \overline{\mathbb{E}}^{\sigma}(\|\rho_t-\sigma_t\|) \leq (|\alpha|+\gamma)^{\delta}, \quad \forall t\in[0,T_{A,B}(\alpha,\gamma)],
    $$
    where $T_{A,B}(\alpha,\gamma):=\frac{1}{B}\log\big(1+\frac{1}{A(|\alpha|+\gamma)^{1-\delta}}\big)$.
\end{proposition}

\proof 
Denote $\Delta_t:=\rho_t-\sigma_t$ and $\hat{\Delta}_t:=\hat{\rho}_t-\hat{\sigma}_t$. By It\^o's formula, we have
\begin{align*}
   \overline{\mathbb{E}}^{\sigma}(\|\Delta_t\|^2)=\overline{\mathbb{E}}^{\sigma}\int^t_0 2\Tr\Big[&\Delta_s\Big(S(\rho_s,\hat{\rho}_s)-S(\sigma_s,\hat{\sigma}_s)+\sum^n_{k=1}\theta_k\big(\mathcal{D}_{L_k}(\rho_s)-\mathcal{D}_{L_k}(\sigma_s)\big)\\
&-F_{\alpha,\gamma}(\sigma_s) \Big) \Big]ds+\overline{\mathbb{E}}^{\sigma}\int^t_0\sum^n_{k=1}\eta_k\theta_k\|\mathcal{G}_{L_k}(\rho_s)-\mathcal{G}_{L_k}(\sigma_s)\|^2ds,
\end{align*}
where $S(\rho,\hat{\rho}):=[-i(H_0+u(\hat{\rho})H_1),\rho]$, and 
\begin{align*}
   \overline{\mathbb{E}}^{\sigma}(\|\hat{\Delta}_t\|^2)=&\overline{\mathbb{E}}^{\sigma}\int^t_0 2\Tr\Big[\hat{\Delta}_s\Big(S(\hat{\rho}_s,\hat{\rho}_s)-S(\hat{\sigma}_s,\hat{\sigma}_s)+\sum^n_{k=1}\hat{\theta}_k\big(\mathcal{D}_{L_k}(\hat{\rho}_s)-\mathcal{D}_{L_k}(\hat{\sigma}_s)\big)\\
   &~~~~~~~~~~~~~~+\sum^n_{k=1} \sqrt{\hat{\eta}_k\hat{\theta}_k}\big(\mathcal{G}_{L_k}(\hat{\rho}_s)\mathcal{T}_k(\rho_t,\hat{\rho}_t)-\mathcal{G}_{L_k}(\hat{\sigma}_s)\mathcal{T}_k(\sigma_t,\hat{\sigma}_t)\big) \Big) \Big]ds\\
   &~~~~~~~~~~~~~~+\overline{\mathbb{E}}^{\sigma}\int^t_0\sum^n_{k=1}\hat{\eta}_k\hat{\theta}_k\|\mathcal{G}_{L_k}(\rho_s)-\mathcal{G}_{L_k}(\sigma_s)\|^2ds.
\end{align*}
Due to the Lipschitz continuity of $S(\rho,\hat{\rho})$, there exists $c_1>0$ such that $\|S(\rho,\hat{\rho})-S(\hat{\sigma},\sigma)\|\leq c_1(\|\Delta\|+\|\hat{\Delta}\|)$. By Cauchy-Schwarz inequality, there exists $c_2>0$ such that $\Tr\big[\Delta\big(S(\rho,\hat{\rho})-S(\hat{\sigma},\sigma)\big)\big]\leq c_2(\|\Delta\|^2+\|\Delta\|\|\hat{\Delta}\|)$. By similar arguments, there exist $c_3,c_4,c_5>0$ such that $\Tr\big[\Delta\big(\mathcal{D}_{L_k}(\rho_s)-\mathcal{D}_{L_k}(\sigma_s) \big)\big]\leq c_3\|\Delta\|^2$, $\Tr\big(\Delta F_{\alpha,\gamma}(\sigma)\big)\leq c_4(|\alpha|+\gamma)  \|\Delta\|$ and $\|\mathcal{G}_{L_k}(\rho_s)-\mathcal{G}_{L_k}(\sigma_s)\|^2\leq c_5\|\Delta\|^2$. Thus, there are three constants $c_6,c_7,c_8>0$ such that
\begin{align*}
   \overline{\mathbb{E}}^{\sigma}(\|\Delta_t\|^2)\leq \int^t_0 c_6\overline{\mathbb{E}}^{\sigma}(\|\Delta_s\|^2)+ c_7(\alpha+\gamma)\overline{\mathbb{E}}^{\sigma}(\|\Delta_s\|)+c_8\overline{\mathbb{E}}^{\sigma}(\|\Delta_s\|\|\hat{\Delta}_s\|)ds .
\end{align*}
Similarly, we can obtain the following estimation for $\hat{\Delta}_t$,
\begin{align*}
   \overline{\mathbb{E}}^{\sigma}(\|\hat{\Delta}_t\|^2)\leq \int^t_0 \hat{c}_1\overline{\mathbb{E}}^{\sigma}(\|\hat{\Delta}_s\|^2)+\hat{c}_2\overline{\mathbb{E}}^{\sigma}(\|\Delta_s\|\|\hat{\Delta}_s\|)ds
\end{align*}
for some $\hat{c}_1,\hat{c}_2>0$. 

Define $U_t:=\overline{\mathbb{E}}^{\sigma}(\|\Delta_t\|^2+\|\hat{\Delta}_t\|^2)$, by Jensen's inequality, 
\begin{align*}
\overline{\mathbb{E}}^{\sigma}(\|\Delta\|)\leq \overline{\mathbb{E}}^{\sigma}(\|\Delta\|)+\overline{\mathbb{E}}^{\sigma}(\|\hat{\Delta}\|)\leq \sqrt{\overline{\mathbb{E}}^{\sigma}(\|\Delta\|^2)}+\sqrt{\overline{\mathbb{E}}^{\sigma}(\|\hat{\Delta}\|^2)}\leq \sqrt{2U},
\end{align*}
where we used the fact $\sqrt{x}+\sqrt{y}\leq \sqrt{2(x+y)}$ for $x,y>0$ in the last inequality. Due to $xy\leq (x^2+y^2)/2$, there exist two constants $a,b>0$ such that
$$
U_t\leq \int^t_0 a U_s+b(|\alpha| + \gamma) \sqrt{U_s}ds,
$$
by applying the generalized Gr\"onwall inequality~\cite[pp. 360-361]{mitrinovic1991inequalities}, we have 
$$
U_t\leq\Big( \frac{b(|\alpha| + \gamma)}{2}\int^t_0 e^{a(t-s)/2}ds\Big)^2, 
$$
which implies $\sqrt{2U_t}\leq (|\alpha| + \gamma) A (e^{Bt}-1)$ for some constants $A,B>0$. 
Moreover, for any $\delta\in(0,1)$, we have
$$
(|\alpha| + \gamma)^{1-\delta} A (e^{Bt}-1)\leq 1, \quad t\in[0,T_{A,B}(\alpha,\gamma)],
$$
where $T_{A,B}(\alpha,\gamma)$ goes to infinity when $\alpha$ and $\gamma$ tends to zero. Hence, we have 
$$
\overline{\mathbb{E}}^{\sigma}(\|\Delta_t\|)\leq \sqrt{2U_t}\leq (|\alpha|+\gamma)^{\delta}, \quad \forall t\in[0,T_{A,B}(\alpha,\gamma)],
$$ 
that completes the proof.
\hfill$\square$

\begin{remark}
     Proposition~\ref{Prop:Stability} examines the average difference between the trajectories of the nominal system $\rho_t$, and that of the system under general perturbation $\sigma_t$, within a finite time horizon. Furthermore, assuming that $\eta_k < 1$ for all $k \in [n]$, and given that the assumptions \textbf{\emph{H1}}, \textbf{\emph{H2}}, \textbf{\emph{A1.1}}-\textbf{\emph{A2}}, as well as the conditions \textbf{\emph{C1}} and \textbf{\emph{C2}} are met, Theorem~\ref{Thm:GES Feedback} establishes that $\rho_t$ converges exponentially to $\mathcal{I}(\mathcal{H}_S)$ with a Lyapunov exponent bounded above by $-\mathsf{C}-\mathsf{K}$, $\overline{\mathbb{P}}^{\sigma}$-almost surely. Specifically, for any initial state $\sigma \in \mathcal{S}(\mathcal{H})$, there exists a finite random variable $R$ such that $\mathbf{d}_{0}(\rho_t) \leq R e^{-(\mathsf{C}+\mathsf{K})t}$ for all $t \geq 0$, $\overline{\mathbb{P}}^{\sigma}$-a.s.  Consequently, it is reasonable to hypothesize that $\mathbf{d}_{0}(\sigma_t)$ may also converge towards a value associated with the target subspace $\mathcal{H}_S$. However, a detailed analysis of how $\sigma_t$ aligns with the target subspace $\mathcal{H}_S$ is beyond the scope of the current discussion and will be addressed in our future work.
\end{remark}

\subsection{Impact of general perturbations on stability in probability}
In the following, we first present a Lyapunov-based approach for analyzing classical stochastic systems, which allows us to investigate how general perturbations affect the stability of the system in probability. 
Specifically, we consider a classical stochastic differential equation and introduce an unknown perturbation in the drift term by adding $\zeta f_{\rm dis}(q)$, where $\zeta\geq 0$,
\begin{equation}
    dq_t = f(q_t)dt + \zeta f_{\rm dis}(q_{t}) dt + g(q_t)dw_t ,
    \label{Eq:SDE-P}
\end{equation}
where $q_t$ takes values in $Q\subset \mathbb{R}^p$ and $w$ is a one-dimensional standard Wiener process.  
Assume that $f$, $g$, and $f_{\rm dis}$ are appropriately defined functions so that $\{ q_{t}\} _{t\geq 0}$ becomes a unique strong regular solution.  

Let $\bar{S}\subset Q$ be a target subset of a control problem. 
Denote $\mathcal{K}$ as the family of all continuous non-decreasing functions $\mu:\mathbb{R}_{\geq 0}\rightarrow\mathbb{R}_{\geq0}$ such that $\mu(0)=0$ and $\mu(r)>0$ for all $r>0$. 
Moreover, we make the following assumption:
\begin{itemize}
    \item[\textbf{H3}]: there exists $V\in\mathcal{C}^2(Q,\mathbb{R}_{\geq 0})$ such that $V(q)=0$ if and only if $q\in\bar{S},$ and a function $\mu\in\mathcal{K}$ such that, 
$
\mathscr{L} V(q) \leq - \mu (V(q))+\zeta D
$    
for all $q \in S_{c}:= \{ q\in Q\setminus \bar{S}|\, V(q)  < c \}$ for some $c,D>0$, where $\mathscr{L}$ is the semi-group generator associated to~\eqref{Eq:SDE-P} defined as 
\begin{align*}
\mathscr{L}V(q):=&\sum_{i=1}^p\frac{\partial V(q)}{\partial q_i}( f_i(q) + \zeta f_{{\rm dis},i}(q_{t}) )
+\frac12 \sum_{i,j=1}^p\frac{\partial^2 V(q)}{\partial q_i\partial q_j}g_i(q)g_j(q).
\end{align*} 
\end{itemize}

The following lemma shows how the unknown $\zeta$ and $f_{\rm dis}$ can deteriorate the stability.  
\begin{lemma} \label{prop:deteriorate_stability}
Assume that \emph{\textbf{H3}} is satisfied.  
For any $\varepsilon \in (0,1)$ and $r \in (0,c)$, there exists $\delta = \delta (\varepsilon,r) \in (0,r)$ such that for all $q_{0} \in S_{\delta}$, 
\begin{align}
     \mathbb{P} &( V(q_t) <r, \,\forall t\geq 0) 
    \geq 
    1 - \varepsilon -\zeta D \mathbb{E}(\tau _{r})/r,
    \label{Eq:deteriorate_stability}
\end{align}  
where $\tau _{r} := \inf \{ t\geq 0|\, V(q_t)=r \}$.
Moreover, the stability in probability is restored when $\zeta$ tends to zero.
\end{lemma}
\textit{Proof.}
The proof basically follows the arguments of \cite[Theorem 4.2.2]{mao2007stochastic}. For any $\varepsilon \in (0,1)$ and $r \in (0,c)$, we can find $\delta = \delta (\varepsilon,r) >0$ such that 
$
\sup _{q \in S_{\delta}} V(q) < \varepsilon r.  
$
Then, It\^o's formula gives 
\begin{align*}
    \mathbb{E}[V(q_{\tau _{r} \wedge t} )] \leq V(q_{0}) 
    - \mathbb{E} \int _{0} ^{\tau _{r} \wedge t} \mu(V(q_s)) ds  + \zeta D \mathbb{E}(\tau _{r} \wedge t),
\end{align*}
where $\tau _{r} \wedge t := \min \{ \tau_r, t\}$.  Using non-negativity of $V$ and the definition of $\mu$, 
\begin{align*}
    \mathbb{E}[V(q_{\tau _{r} \wedge t})]
    \geq &
    \mathbb{E}[
    \mathds{1}_{\{\tau _{r} \leq t\}} V(q_{\tau _{r}}) ]
    \geq 
    \mathbb{P} (\tau _{r} \leq t) r
    .
\end{align*}
Since $\sup _{q_{0} \in S_{\delta}} V(q_{0}) < \varepsilon r$ and $r>0$, 
\begin{align}
    \mathbb{P} (\tau _{r} \leq t)
    \leq 
    \varepsilon + \zeta D \mathbb{E}(\tau _{r} \wedge t)/r.
    \label{Eq:Ineq}
\end{align}
By monotone convergence theorem, 
 $  
    \mathbb{P} (\tau _{r} < \infty) 
    \leq
    \varepsilon + \zeta D \mathbb{E}(\tau _{r})/r.
$    

Then, for the inequality~\eqref{Eq:Ineq}, let $\zeta$ tend to zero, we have $\mathbb{P} (\tau _{r} \leq t) \leq \varepsilon$, which implies that $\bar{S}$ is stable in probability by letting $t\rightarrow\infty$.
\hfill$\square$ 

Next, by using the above lemma, we investigate how perturbations affect the stability of the nominal quantum system ~\eqref{Eq:SME-Pf}--\eqref{Eq:SME-F} in probability.

\begin{proposition}
\label{Prop:general_perturbation_old}
Suppose that there exist a function $\mathsf{V}\in\mathcal{C}^2(\mathcal{S}(\mathcal{H})\times\mathcal{S}(\mathcal{H}),\mathbb{R}_{\geq 0})$, a constant $l>0$ and $\mu\in\mathcal{K}$ such that $\mathscr{L}\mathsf{V}\leq -\mu(\mathsf{V})$ whenever $\mathsf{V}<l$, where $\mathscr{L}$ is associated to the nominal system and filter pair ~\eqref{Eq:SME}--\eqref{Eq:SME-F}. Then, for coupled the perturbed system/filter~\eqref{Eq:SME-P}--\eqref{Eq:SME-F},
for all $\varepsilon>0$ there exist $\delta\in(0,r)$, $c>0$, and $\zeta >0$ such that 
$$
\mathbb{P} \big( \mathsf{V}(\sigma_t,\hat{\sigma}_t) <l, \,\forall t\geq 0\big) \geq 1 - \varepsilon -\zeta c \mathbb{E}(\tau _{l})/l
$$
whenever $\mathsf{V}(\sigma_0,\hat{\sigma}_0)<\delta$, where $\tau_l$ denotes the first exiting time of $(\sigma_t,\hat{\sigma}_t)$ from $\{\mathsf{V}<l\}$. Moreover, the stability in probability is restored when $\zeta$ tends to zero.
\end{proposition}
\textit{Proof.}
Due to the continuity of $\partial \mathsf{V}/\partial \sigma$ and the compactness of $\mathcal{S}(\mathcal{H})$, there exist constants $D:= D(\alpha,\beta ,\gamma )>0$ and $\zeta = \zeta ( \alpha,\beta,\gamma ) \geq 0 $, where $D(0,0,0)>0$ and $\zeta (0,0,0) = 0$, such that $\bar{\mathscr{L}}\mathsf{V}\leq -\mu(\mathsf{V})+\zeta D$ where $\bar{\mathscr{L}}$ is associated to~\eqref{Eq:SME-P}--\eqref{Eq:SME-F}. The result can be concluded by applying Lemma~\ref{prop:deteriorate_stability}.
\hfill$\square$

The approach of \cite{liang2021robust} can be used in order to find a nominal system/filter that admits a Lyapunov function $\mathsf{V}$ as in the above proposition so that the latter can be specialized to the case of feedback-controlled QSMEs. 
We impose the following condition to ensure that \textbf{H3} is satisfied for the perturbed system~\eqref{Eq:SME-Pf}--\eqref{Eq:SME-F}, especially, it guarantees the local stability in probability of the estimator~\eqref{Eq:SME-F} with respect to the target subspace $\mathcal{H}_S$,
\begin{itemize}
    \item[\textbf{C2'}:]  $\min_{k\in[n]}\big\{\hat{\eta}_k\hat{\theta}_k\underline{\mathfrak{l}}_k^2\big\}>4\mathfrak{C}$.
\end{itemize}
\begin{proposition}
\label{Prop:general_perturbation}
Suppose that the assumptions \emph{\textbf{H1}}, \emph{\textbf{A1.1}}-\emph{\textbf{A1.3}} as well as the conditions \emph{\textbf{C1}} and \emph{\textbf{C2'}} are satisfied. Then, for the perturbed system~\eqref{Eq:SME-Pf}--\eqref{Eq:SME-F},  for all $\varepsilon>0$ there exist $\delta,l>0$ such that 
$$
\overline{\mathbb{P}}^{\sigma} \big( \mathbf{d}_0(\sigma_t) <l, \,\forall t\geq 0\big) \geq 1 - \varepsilon -2(|\alpha|+\gamma)\overline{\mathbb{E}}^{\sigma}(\tau_{l})/l
$$
whenever the initial condition satisfy $(\sigma,\hat{\sigma})\in B_{\delta}(\mathcal{H}_S)\times \mathrm{int}\{\mathcal{S}(\mathcal{H})\}\cap B_{\delta}(\mathcal{H}_S)$, where $\tau_l$ denotes the first exiting time of $(\sigma_t,\hat{\sigma}_t)$ from $B_{l}(\mathcal{H}_S)\times B_{l}(\mathcal{H}_S)$. Moreover, the stability in probability is restored when $\alpha$ and $\gamma$ tend to zero.
\end{proposition}
\textit{Proof.}
Consider the function $\mathsf{V}(\sigma,\hat{\sigma}):=1-\Tr(\Pi_0\sigma)+\sum^d_{j=1}\sqrt{\mathrm{Tr}(\hat{\sigma} \Pi_j)}\geq 0$, where equality holds if and only if $(\sigma,\hat{\sigma})\in \mathcal{I}(\mathcal{H}_S)\times \mathcal{I}(\mathcal{H}_S)$. By the similar arguments as in the proof of Theorem~\ref{Thm:GES Feedback}, for all $(\sigma,\hat{\sigma})\in B_{\varepsilon}(\mathcal{H}_S)\times \mathrm{int}\{\mathcal{S}(\mathcal{H})\}\cap B_{\varepsilon}(\mathcal{H}_S)$, we have
$\mathscr{L}\mathsf{V}(\sigma,\hat{\sigma})\leq -\mathbf{c}(\sigma,\hat{\sigma})\sum^d_{j=1}\sqrt{\mathrm{Tr}(\hat{\sigma} \Pi_j)}+|\Tr(\Pi_0 F_{\alpha,\gamma}(\sigma))|$,
where 
$$
\mathbf{c}(\sigma,\hat{\sigma}):=\frac{1}{2}\min_{k\in[n]}\big\{\hat{\eta}_k\hat{\theta}_k(\underline{\mathfrak{l}}_k-|\mathbf{P}_{k,0}(\hat{\sigma})|)^2\big\}-
    \sum^n_{k=1}\sqrt{\hat{\eta}_k\hat{\theta}_k}(\bar{\mathfrak{l}}_k+|\mathbf{P}_{k,0}(\hat{\sigma})|)|\mathcal{T}_k(\sigma,\hat{\sigma})|.
$$
By a straightforward computation, we have $|\Tr(\Pi_0 F_{\alpha,\gamma}(\sigma))|\leq 2(|\alpha|+\gamma)$ for all $\sigma\in\mathcal{S}(\mathcal{H})$, and 
$\lim_{(\sigma,\hat{\sigma})\rightarrow \mathcal{I}(\mathcal{H}_S)\times \mathcal{I}(\mathcal{H}_S)} \mathbf{c}(\sigma,\hat{\sigma}) = \mathsf{c} > 0,$
where $\mathsf{c}= \frac{1}{2}\min_{k\in[n]}\big\{\hat{\eta}_k\hat{\theta}_k\underline{\mathfrak{l}}_k^2\big\}-2\mathfrak{C}$ and the positivity is ensured by \textbf{C2'}. Thus, there exist $c\in(0,\mathsf{c})$ and $l\in(0,\kappa)$, where $\kappa>0$ is defined in \textbf{H1}, such that 
$$
\mathscr{L}\mathsf{V}(\sigma,\hat{\sigma}) \leq -c\sum^d_{j=1}\sqrt{\mathrm{Tr}(\hat{\sigma} \Pi_j)}+2(|\alpha|+\gamma), \quad \forall (\sigma,\hat{\sigma})\in B_{l}(\mathcal{H}_S)\times \mathrm{int}\{\mathcal{S}(\mathcal{H})\}\cap B_{l}(\mathcal{H}_S).
$$
The result can be concluded by applying Lemma~\ref{prop:deteriorate_stability} and Lemma~\ref{Lemma:PosDef invariant}, along with the relation $\mathbf{d}_0(\sigma)^2\leq 3(1-\Tr(\Pi_0\sigma))$ which is established in the proof of Proposition~\ref{Prop:ISS}.
\hfill$\square$

\section{Conclusions}
\label{Sec:Conclusion}

We have analyzed four scenarios in which perturbations enter a QSME that globally stabilizes a target subspace: we consider open-loop and feedback control, in combination with perturbations that are either invariance preserving or not.
The first general conclusion is that invariance is critical: if the perturbations leave the target invariant, under reasonable conditions one can show that stability is also preserved. In addition, in order to prove these types of results for feedback systems, it is important to have accurate knowledge of the measurement operators (see the introduction of Section \ref{Sec:Feedback}).
For general perturbations, we were able to provide bounds on the effect of the perturbation in probability or in mean. Stronger results may be derived by leveraging ergodic properties of the solutions, which we leave for future developments of this research line.
We believe this work represents a first step towards a systematic analysis of the robustness of stabilizing controls for quantum systems, providing useful indications and bounds on the critical parameters to be designed or characterized before hands.

\appendix

{
\section{Dissipation-Induced Decomposition essentials}
\label{Sec:DID}
A method to decide if a subspace is attracting has been proposed in \cite{ticozzi2012hamiltonian}, which is based on a decomposition of the Hilbert space that also allows one to study the speed of convergence towards the target and related geometrical properties (see also \cite{cirillo} for a discrete-time version). We briefly recall here some basic definitions and facts, which are instrumental to proving Theorem \ref{thm:genericgas} below, and thus to prove that perturbations that preserve invariance will generically also preserve GAS. 

Let $\Hi_S$ be a proper subspace of $\Hi.$ Then it can be proved that $ \Hi_S$ is GAS for the semigroup dynamics $$\dot\rho=-i[H,\rho]+\sum_k L_k\rho L_k^* -\frac{1}{2}\{L_k^* L_k,\rho\}$$ if
and only if a Hilbert space decomposition in orthogonal subspaces, of
the form \begin{equation}
\label{DID}
\Hi=\Hi_S\oplus\Hi^{(1)}_T\oplus\Hi^{(2)}_T\ldots\oplus\Hi^{(q)}_T,\end{equation}
can be obtained by the algorithm presented in detail in the next section  \cite{ticozzi2012hamiltonian}. Such
decomposition is called the {\em Dissipation-Induced Decomposition}
(DID).  Each of the subspaces $\Hi^{(i)}_T$ in the direct sum is
referred to as a {\em basin}, and is associated to an active dynamical connection to the preceding one.

Partitioning each matrix in
blocks according to the subspaces generated by the DID leads to a standard
structure, where the upper block-diagonal blocks establish the
dissipation-induced, cascade connections between the different basins
$\Hi_T^{(i)}:$
\begin{eqnarray}
L_k=
\left[\begin{array}{c|cccc}  
L_S & \hat L_P^{(0)} & 0 & \cdots &  \\
\hline	0 &  L_T^{(1)} & \hat L_P^{(1)} & 0 & \cdots\\
\vdots & L_Q^{(1)} & L_T^{(2)} & \hat L_P^{(2)} & \ddots\\
 &  \vdots & \ddots & \ddots & \ddots\\
\end{array}\right]_k .
\label{matDID}
\end{eqnarray}
Similarly, the Hamiltonian becomes:
\begin{eqnarray}
H =
\left[\begin{array}{c|cccc}  
H_S &  H_P^{(0)} & 0 & \cdots &  \\
\hline	H_P^{(0)\dag} &  H_T^{(1)} & \cdots &  & \\
0 & \vdots & \ddots &  & \\
\vdots &   &  &  & \
\end{array}\right]_k . 
\label{matHDID}
\end{eqnarray}
By construction, the $\hat L_P^{(i)}$ blocks are either zero or full
rank. The fact that the first column of blocks has only $L_S\neq 0$ is
a necessary condition for the invariance of $\Hi_S.$ It
follows that $\hat L_P^{(0)}\neq 0,$ otherwise $\Hi_S$
cannot be GAS.

In order to make the presentation self-contained, we reproduce here the
algorithm for the construction of the DID \cite{ticozzi2012hamiltonian}.  
The inputs are a Lindblad generator ${\cal L}$ associated to Hamiltonian
$H$ and noise operators $\{L_k\},$  and the target invariant subspace $\Hi_S$.  Checking invariance can be 
done using Lemma \ref{Lem: Invariant}.

\noindent \rule{\linewidth}{1pt} \\[-1mm] {\em Algorithm for GAS
verification and DID construction} \\ [-2.5mm]
\noindent \rule{\linewidth}{1pt}  

Let $\Hi_S$ be invariant. Call $\Hi_R^{(0)}:=\Hi_R,$
$\Hi_S^{(0)}:=\Hi_S,$ choose an orthonormal basis for the subspaces
and write the matrices with respect to that basis. Rename the matrix
blocks as follows: $H_{S}^{(0)}:= H_{S},$ $H_{P}^{(0)}:= H_{P},$
$H_{R}^{(0)}:=H_{R},$ $L_{S,k}^{(0)}:=L_{S,k},$
$L_{P,k}^{(0)}:=L_{P,k}$, and $L_{R,k}^{(0)}:=L_{R,k}.$ 

For $j\geq 0$, consider the following iterative procedure:

\begin{enumerate}
\item Compute the matrix blocks $L_{P,k}^{(j)}$ according to the
decomposition ${\Hi}^{(j)}=\Hi_S^{(j)}\oplus\Hi_R^{(j)}.$
\item Define $ \Hi_R^{(j+1)}:=\bigcap_k\ker L_{P,k}^{(j)}.$
\item Consider the following three sub-cases:
\begin{itemize}
\item[a.] If $ \Hi_R^{(j+1)}=\{0\}$,  define $\Hi_T^{(j+1)}:=\Hi^{(j)}_R.$ 
The iterative procedure is successfully completed.

\item[b.] If $ \Hi_R^{(j+1)}\neq\{0\},$ but $ \Hi_R^{(j+1)}\subsetneq
\Hi_R^{(j)},$ define $\Hi_T^{(j+1)}$ as the orthogonal complement of
$\Hi_R^{(j+1)}$ in $\Hi_R^{(j)}$, that is,
$\Hi_R^{(j+1)}=\Hi_R^{(j)}\ominus\Hi_R^{(j+1)}.$ 

\item[c.] If  $ \Hi_R^{(j+1)}= \Hi_R^{(j)}$ (that is, $L_{P,k}^{(j)} =
0 \; \forall k$), define
$$\tilde{\cal L}_P^{(j)}:=-i {H}_P^{(j)}-\frac{1}{2}\sum_k
{L}_{Q,k}^{(j)\dag}L_{R,k}^{(j)}.$$
\begin{itemize}
\item If $\tilde{\cal L}_P^{(j)}\neq0,$ re-define
$\Hi_R^{(j+1)}:=\ker(\tilde{\cal L}^{(j)}_P)$.\\ If $
\Hi_R^{(j+1)}=\{0\}$, define $\Hi_T^{(j+1)}:=\Hi^{(j)}_R$ and the iterative procedure is successfully completed. Otherwise, define
$\Hi_T^{(j+1)}:=\Hi_R^{(j)}\ominus\Hi_R^{(j+1)}$.

\item If $\tilde{\cal L}^{(j)}_P=0,$ then $\Hi_R^{(j)}$ is invariant
and $\Hi_S$ cannot be GAS. Exit the algorithm.
\end{itemize}

\end{itemize}

\item Define $\Hi_S^{(j+1)}:=\Hi_S^{(j)}\oplus \Hi^{(j+1)}_T.$ To
construct a basis for $\Hi_S^{(j+1)},$ append to the {\em already
defined} basis for $\Hi^{(j)}_S$ an orthonormal basis for $
\Hi^{(j+1)}_T.$

\item Increment the counter $j$ and go back to step 1.
\end{enumerate}

\noindent \rule{\linewidth}{1pt} \\

\vspace*{1mm}

\noindent 
The algorithm ends in a finite number of steps, since at every
iteration it either stops or the dimension of $\Hi^{(j)}_R$ is reduced
by at least one.

\section{GAS is mean is generic under invariance}

The following results extend some of the results presented in \cite{ticozzi2014qls}, and uses the known as DID decomposition, introduced in \cite{ticozzi2012hamiltonian}, to prove recursively that once invariance is guaranteed, asymptotic stability is generic when achievable.

\begin{theorem}\label{thm:genericgas}
     Consider a system in Lindblad form:
     \begin{equation}\label{eq:lindmu}
         \dot\rho=-i[H(x),\rho]+\textstyle\sum_k{\cal D}_{L_k(x)}(\rho),
     \end{equation}
     where $x\in\mathbb{R}^n$ is a set of parameters and the dependence of $H$ and $C_k$ on the parameters is polynomial. Assume that ${\cal H}_S$ is invariant for  \eqref{eq:lindmu} for any $x.$ Then if there is a choice $\bar x$ such that $\mathcal{H}_S$ is GAS then it is GAS also for almost all choices of $x$.
\end{theorem}
\textit{Proof.}
Define an $m\times n$ matrix $X=[f_{jk}(x)],$ with
$f_{jk}:\R^K\rightarrow\C$, such that the real and imaginary parts
$\Re(f_{jk}),\Im(f_{jk})$ are (real)-analytic, and let $r_m \equiv
\max_{x\in\C^K}\text{rank}(X).$ Notice that
$\text{rank}(X)\in\{0,\ldots,\text{min}\{n,m\}\}$ for all $x\in \C^K,$ and
the maximum is attained in $\R^K$. 
Then the set ${\cal X}=\{x\in \R^K\,|\,\textrm(rank)(X)<r_m\}$ is such that
$\mu({\cal X})=0,$ see e.g.  \cite{ticozzi2014qls}, Lemma A.2.
 
We now focus on the DID that makes $\rho_d$ GAS, and  prove
it is actually generic, leveraging this observation. By construction, the matrix block decomposition of the matrix
representation of $H(\bar x),\{L_k(\bar x)\}$ must be such that for each iteration,
indexed by $j,$ either \begin{equation}
    \tilde{\cal D}^{(j)}:=
\left[\begin{array}{c}L_{P,1}^{(j)}\\\vdots \\
L_{P,M}^{(j)}\end{array}\right]\label{rp1}\end{equation}
\noindent 
has {\em maximum rank}
$r^{(j)}=\max\{\dim(\Hi_T^{(j-1)}),\dim(\Hi_T^{(j)})\}$ (see Step 2, 
3.a and 3.b), or \begin{equation}
    \tilde{\cal L}^{(j)}_P:=i
{H}_P^{(j)}-\frac{1}{2}\textstyle\sum_k
{L}_{Q,k}^{(j)\dag}L_{T,k}^{(j)}\label{rp2}\end{equation}
\noindent 
has {\em full rank} $r^{(j)}$ (defined as above, see step 3.c of the
algorithm).  At the $j$-th iteration, $\dim(\Hi_T^{(j-1)})$ is fixed,
but almost all choices of parameters
 provide a maximal $\dim(\Hi_T^{(j)})$: in fact,
given Eq. \eqref{rp1}-\eqref{rp2}, this is equivalent to maximize the
rank of either $\tilde{\cal D}^{(j)}$ or $\tilde{\cal L}^{(j)}_P$ at
each iteration.  As the elements of $\tilde{\cal D}^{(j)}$ and
$\tilde{\cal L}^{(j)}_P$ are (complex) polynomial functions of the
real parameters $x,$  we have that the parameter set corresponding to
the maximal rank of the corresponding matrices has measure $1.$ If $\bar x$ maximizes the rank, then almost every other choice of $x$ will also lead do so. Hence the same idea of the proof showing that the complete DID leads to a GAS subspace can be applied to the perturbed generator. 
In fact, for $\bar x$ we obtain a decomposition $\Hi=\Hi_S\oplus\Hi_R$ where $\Hi_R=\Hi^{(1)}_T\oplus\Hi^{(2)}_T\oplus\ldots\oplus\Hi^{(q)}_T.$ 
We can prove by (finite) induction that no invariant subspace is
contained in $\Hi_R$ also for almost all $x.$

Consider the last subspace $\Hi^{(q)}_T.$ By construction for $\bar x$ either $\bigcap_k
\ker (L_{P,k}^{(q)}) =\{0\},$ or $L_{P,k}^{(q)}=0$ and $\tilde{\cal L}^{(q)}_P$ is
full column-rank. We recalled at the beginning that the same must be true for almost all $x.$ In either case, $\Hi^{(q)}_T$ cannot contain any invariant set since the dynamics drives any state with support only in $\Hi^{(q)}_T$ out of the subspace, which cannot thus contain any invariant set \cite{ticozzi2008quantum,ticozzi2012hamiltonian}.

Next, working by backward finite induction, we assume that
$\Hi^{(\ell+1)}_T\oplus\ldots\oplus\Hi_T^{(q)}$, $\ell+1 \leq q$, does not
contain invariant subspaces, while (by contradiction)
$\Hi^{(\ell)}_T\oplus\Hi^{(\ell+1)}_T\oplus\ldots\oplus\Hi^{(q)}_T$
does. Then the invariant subspace should be non-orthogonal to
$\Hi^{(\ell)}_T,$ which is in the algorithm defined as the orthogonal complement
of either $\bigcap_k \ker (L_{P,k}^{(\ell -1)})$  
or $\ker (\tilde{\cal L}_P^{(\ell-1)}).$ But then any state $\rho$
on
$\Hi^{(\ell)}_T\oplus\Hi^{(\ell+1)}_T\oplus\ldots\oplus\Hi^{(q)}_T$
and non-trivial support on $\Hi^{(\ell)}_T$ would violate the
invariance conditions by the same argument of the first step ($\Hi^{(q)}$ above).  By iterating until $\ell=1$, we have that $\Hi_R$
cannot contain invariant subspaces for almost all $x$, and thus $\Hi_S$ is GAS \cite{ticozzi2008quantum}. }
\hfill$\square$

\section{Recurrence of the system with feedback under invariance}
\label{Sec:Recurrence}
Firstly, we state some fundamental results that are used in the proof of instability and recurrence. These results are analogous to the results in~\cite[Section~4]{liang2019exponential} and \cite[Lemma 7]{liang2021GHZ} for the coupled system~\eqref{Eq:SME-Pf}--\eqref{Eq:SME-F}, and they concern invariance properties for the system. Since their proofs are based on the same arguments, we omit them. 
\begin{lemma}
Assume that \emph{\textbf{H1}} holds. The ranks of ${\sigma}_t$ and $\hat{\sigma}_t$ are $\overline{\mathbb{P}}^{\sigma}$-almost surely non-decreasing.
\label{Lemma:PosDef invariant}
\end{lemma}

\begin{lemma}\label{Lemma:Mixed}
   Assume that \emph{\textbf{H1}} and \emph{\textbf{H2}} are satisfied. In addition, suppose that $\eta_k\in(0,1)$ for all $k\in[n]$. Then, for all initial state $\sigma_0=\sigma\in\{\rho\in\mathcal{S}|\,\Tr(\rho^2)=1\}\setminus \mathcal{I}(\mathcal{H}_S)$, $\sigma_t$ is mixed (i.e., $\Tr(\sigma_t^2)<1$) for all $t>0$, $\overline{\mathbb{P}}^{\sigma}$-almost surely.   
\end{lemma}

Secondly, we explore the recurrence property of the trajectories of the coupled system~\eqref{Eq:SME-Pf}--\eqref{Eq:SME-F}.
It is obvious to verify that, under the assumption \textbf{H1} and \textbf{H2}, the estimator~\eqref{Eq:SME-F} contains only one invariant subspace $\mathcal{H}_S$. Then, we consider the case $u\equiv 0$, since we have only limited information on the perturbations, for the perturbed system~\eqref{Eq:SME-Pf} under the assumption \textbf{AR'}, there are three possibilities on the invariant subset which depends on the structure of the $\tilde{H_0}$ and $(C_j)_{j\in[m]}$:
\begin{enumerate}
    \item for all $\alpha,\gamma$, the system contains only one invariant subset $\mathcal{I}(\mathcal{H}_S)$;
    \item for all $\alpha,\gamma$, there exists a non-empty subset $E\subset [d]$ such that $\mathcal{I}(\mathcal{H}_S)$ and $\mathcal{I}(\mathcal{H}^j_{R})$ for all $j\in E$ is invariant,
    \item for zero measure set of $\alpha,\gamma$, the system contains several equilibria besides the invariant set motioned in Case 2. 
\end{enumerate}
Note that, if $\tilde{H}_0,C_j\in \mathrm{span}\{\Pi_0,\dots\Pi_d\}$ then Case 2 happens. In the Case 2, for the coupled system~\eqref{Eq:SME-Pf}--\eqref{Eq:SME-F} under the assumptions \textbf{H1}, \textbf{H2} and \textbf{AR'}, there are $\mathrm{card}(E)$ non-desired invariant subsets, $\mathcal{I}(\mathcal{H}^j_R)\times\mathcal{I}(\mathcal{H}_S)$ for all $j\in E$,which cannot be cancelled by the feedback controller $u(\hat{\sigma}_t)$.
Then, we focus on the Case 3, in the following lemma, we provide the sufficient conditions to ensure the instability of the non-desired invariant subsets.
\begin{lemma}
\label{Lem:Instability}
    Suppose that the assumptions \emph{\textbf{AR'}}, \emph{\textbf{H1}}, \emph{\textbf{A1.2}}, \emph{\textbf{A1.3}} and \emph{\textbf{C1}} are satisfied. Consider the case in which there exists a non-empty subset $E\subset [d]$ such that $\mathcal{I}(\mathcal{H}^j_R)$ for all $j\in E$ are also invariant for the perturbed system~\eqref{Eq:SME-Pf} when $u\equiv 0$. Then, for almost all values of $\alpha,\gamma$, there exists $\lambda>0$ such that for all initial condition $(\sigma,\hat{\sigma})\in B_{\lambda}(\mathcal{H}^j_R)\times B_{\lambda}(\mathcal{H}_S)\cap \mathrm{int}\{\mathcal{S}(\mathcal{H})\}$ with $j\in E$, the trajectories $(\sigma_t,\hat{\sigma}_t)$ of the coupled system~\eqref{Eq:SME-Pf}--\eqref{Eq:SME-F} exits $B_{\lambda}(\mathcal{H}^j_R)\times B_{\lambda}(\mathcal{H}_S)$ in finite time $\overline{\mathbb{P}}^{\sigma}$-almost surely.
\end{lemma}
 \proof 
Consider the function $-\log \Tr(\Pi_j \hat{\sigma})$ with $j\in E$, whose infinitesimal generator related to~\eqref{Eq:SME-Pf}--\eqref{Eq:SME-F} is given by
\begin{equation*}
    \mathscr{L} -\log \Tr(\Pi_j \hat{\sigma}) = -\frac{ \mathscr{L}\Tr(\Pi_j \hat{\sigma})}{\Tr(\Pi_j \hat{\sigma})}+\frac{\sum^n_{k=1}\hat{\eta}_k\hat{\theta}_k\Tr\big(\Pi_j\mathcal{G}_{L_{k}}(\hat{\sigma})\big)^2}{2\Tr(\Pi_j\hat{\sigma})^2}.
\end{equation*}
For all $\hat{\sigma}\in B_{\kappa}(\mathcal{H}_S)$ where $\kappa>0$ is defined in \textbf{H1}, we have $u(\hat{\sigma})=0$ due to \textbf{H1}. Together with \textbf{AR'}, we deduce  
\begin{equation*}
    \frac{ \mathscr{L}\Tr(\Pi_j \hat{\sigma})}{\Tr(\Pi_j \hat{\sigma})} =2\sum^n_{k=1} \sqrt{\hat{\eta}_k\hat{\theta}_k} \mathcal{T}_k(\sigma,\hat{\sigma})\Big(\mathbf{Re}\{l_{k,j}\}-\sum^{d}_{i=0}\mathbf{Re}\{l_{k,i}\}\Tr(\hat{\sigma}\Pi_i)\Big).
\end{equation*}
It implies 
\begin{align}
    \lim_{(\sigma,\hat{\sigma})\rightarrow \mathcal{I}(\mathcal{H}^j_R)\times \mathcal{I}(\mathcal{H}_S)}& \frac{ \mathscr{L}\Tr(\Pi_j \hat{\sigma})}{\Tr(\Pi_j \hat{\sigma})}\nonumber\\
    &= 4 \sum^n_{k=1}\hat{\eta}_k\hat{\theta}_k \big(\mathbf{Re}\{l_{k,j}\}-\mathbf{Re}\{l_{k,0}\} \big)\big(\chi_k \mathbf{Re}\{l_{k,j}\}-\mathbf{Re}\{l_{k,0}\} \big).\label{Eq:insta1}
\end{align}
Moreover, we have
\begin{align}
     \lim_{(\sigma,\hat{\sigma})\rightarrow \mathcal{I}(\mathcal{H}^j_R)\times \mathcal{I}(\mathcal{H}_S)}\frac{\sum^n_{k=1}\hat{\eta}_k\hat{\theta}_k\Tr\big(\mathcal{G}_{L_{k}}(\hat{\sigma})\Pi_j\big)^2}{2\Tr(\hat{\sigma}\Pi)^2}
    = 2 \sum^n_{k=1}\hat{\eta}_k\hat{\theta}_k \big(\mathbf{Re}\{l_{k,j}\}-\mathbf{Re}\{l_{k,0}\} \big)^2.\label{Eq:insta2}
\end{align}

Then, we obtain
\begin{align*}
    &\lim_{(\sigma,\hat{\sigma})\rightarrow \mathcal{I}(\mathcal{H}^j_R)\times \mathcal{I}(\mathcal{H}_S)}\mathscr{L} -\log \Tr(\Pi_j \hat{\sigma})\\
    &~~~~~~~~=-2 \sum^n_{k=1}\hat{\eta}_k\hat{\theta}_k \big(\mathbf{Re}\{l_{k,j}\}-\mathbf{Re}\{l_{k,0}\} \big)\big((2\chi_k-1) \mathbf{Re}\{l_{k,j}\}-\mathbf{Re}\{l_{k,0}\} \big).
\end{align*}
Under \textbf{A1.3} and \textbf{C1}, for all $j\in[d]$, we deduce that, for all $k\in[n]$ such that $\bar{\mathfrak{c}}_k\leq 0$,
\begin{align*}
    &\mathbf{Re}\{l_{k,j}\}\geq \min_{j\in[d]}\mathbf{Re}\{l_{k,j}\} \geq \mathbf{Re}\{l_{k,0}\},\\
    &(2\chi_k-1) \mathbf{Re}\{l_{k,j}\}\geq (2\chi_k-1) \min_{j\in[d]}\mathbf{Re}\{l_{k,j}\} = (2\chi_k-1)(\mathbf{Re}\{l_{k,0}\}-\bar{\mathfrak{c}}_k) >\mathbf{Re}\{l_{k,0}\},
\end{align*}
where $2\chi_k-1>0$ due to \textbf{C1}; and for all $k\in[n]$ such that $\underline{\mathfrak{c}}_k\geq 0$,
\begin{align*}
    &\mathbf{Re}\{l_{k,j}\}\leq \max_{j\in[d]}\mathbf{Re}\{l_{k,j}\} \leq \mathbf{Re}\{l_{k,0}\},\\
    &(2\chi_k-1) \mathbf{Re}\{l_{k,j}\}\leq (2\chi_k-1) \max_{j\in[d]}\mathbf{Re}\{l_{k,j}\} = (2\chi_k-1)(\mathbf{Re}\{l_{k,0}\}-\underline{\mathfrak{c}}_k) <\mathbf{Re}\{l_{k,0}\}.
\end{align*}
Furthermore, due to \textbf{A1.3}, there exists at least one $k\in[n]$ such that $\mathbf{Re}\{l_{k,j}\}\neq \mathbf{Re}\{l_{k,0}\}$, thus we have 
$$
\lim_{(\sigma,\hat{\sigma})\rightarrow \mathcal{I}(\mathcal{H}^j_R)\times \mathcal{I}(\mathcal{H}_S)}\mathscr{L} -\log \Tr(\Pi_j \hat{\sigma})<0.
$$
Therefore, we can conclude that, there exist $\delta>0$ and $\lambda\in(0,\kappa)$ such that 
\begin{equation*}
     \mathscr{L} -\log \Tr(\Pi_j \hat{\sigma})\leq -\delta, \quad \forall (\sigma,\hat{\sigma})\in B_{\lambda}(\mathcal{H}^j_R) \times B_{\lambda}(\mathcal{H}_S)\cap \mathrm{int}\{\mathcal{S}(\mathcal{H})\}.
\end{equation*}

Define $\tau_{\lambda}$ as the first exiting time from $B_{\lambda}(\mathcal{H}^j_R) \times B_{\lambda}(\mathcal{H}_S)$. 
Due to Lemma~\ref{Lemma:PosDef invariant}, for $\hat{\sigma}_0=\hat{\sigma}>0$, $\hat{\sigma}_t>0$ for all $t\geq 0$, $\overline{\mathbb{P}}^{\sigma}$-almost surely, therefore $\Tr(\Pi_j \hat{\sigma}_t)>0$, $\overline{\mathbb{P}}^{\sigma}$-almost surely.  Thus, we can apply the It\^o's formula on $-\log \Tr(\Pi_j \hat{\sigma}_t)$, which implies $\overline{\mathbb{E}}^{\sigma}(\tau_{\lambda})\leq -\log \Tr(\Pi_j\hat{\sigma}_0)/\delta<\infty$. Then, we can conclude the proof by applying the Markov inequality.
\hfill$\square$

\begin{remark}
\label{Rem:instability}
 We demonstrate the instability of $\mathcal{I}(\mathcal{H}^j_R) \times \mathcal{I}(\mathcal{H}_S)$ by establishing that $\Tr(\Pi_j \hat{\sigma})$ deviates from zero. This deviation indirectly implies that $\Tr(\Pi_0 \hat{\sigma})$ moves away from one as $(\sigma_0,\hat{\sigma}_0)$ approaches the invariant subset $\mathcal{I}(\mathcal{H}^j_R) \times \mathcal{I}(\mathcal{H}_S)$. Assumption \textbf{A1.3} ensures that the sign of $\mathbf{P}_{k,0}(\hat{\sigma})$—and consequently, the sign of the infinitesimal generator of $\Tr(\Pi_0 \hat{\sigma})$—remains constant in a neighborhood of $\mathcal{I}(\mathcal{H}_S)$. Together with condition \textbf{\emph{C1}}, this allows us to assert that the infinitesimal generator of $-\log\Tr(\Pi \hat{\sigma})$ is less than a negative constant near the invariant subset, analogous to the conditions in Khas'minskii's recurrence theorem. Furthermore, in the context of an $N$-level spin system~\cite{liang2019exponential,liang2021robust}, the satisfaction of \textbf{\emph{A1.3}} aligns with scenarios where the states $\boldsymbol{\rho}_0$ or $\boldsymbol{\rho}_{2J}$ are designated as the target states.
\end{remark}

Next, inspired by the arguments in~\cite[Section 4.2]{liang2020robustness}, in the following proposition, we show the recurrence of $(\sigma_t,\hat{\sigma}_t)$ relative to any neighbourhood of $\mathcal{I}(\mathcal{H}_S)\times \mathcal{I}(\mathcal{H}_S)$. Based on the support theorem~\cite{stroock1972support}, the deterministic control system corresponding to the Stratonovich form of the coupled system~\eqref{Eq:SME-Pf}--\eqref{Eq:SME-F} is given by 
\begin{align}
    \dot{\sigma}_v(t) &= \mathfrak{L}^u_{\theta,\eta}(\sigma_v(t))+F_{\alpha,\gamma}(\sigma_v(t))+\textstyle\sum^n_{k=1}\sqrt{\eta_k \theta_k} \mathcal{G}_{L_k}(\sigma_v(t))V_k(t),\label{Eq:ODE-P}\\
    \dot{\hat{\sigma}}_v(t) &= \mathfrak{L}^u_{\hat{\theta},\hat{\eta}}(\hat{\sigma}_v(t))+\textstyle\sum^n_{k=1}\sqrt{\hat{\eta}_k \hat{\theta}_k} \mathcal{G}_{L_k}(\hat{\sigma}_v(t))V_k(t),\label{Eq:ODE-F}
\end{align}
where $\sigma_v(0)=\sigma_0=\sigma$ and $\hat{\sigma}_v(0)=\hat{\sigma}_0=\hat{\sigma}$, $V_k(t):=v_k(t)+\sqrt{\eta_k \gamma_k}\mathrm{Tr}((L_k+L_k^*)\sigma_v(t))$ where $v_k(t)\in\mathcal{V}$ is the bounded control input, where $\mathcal{V}$ is the set of all locally bounded measurable functions from $\mathbb{R}_+$ to $\mathbb{R}$. Here
\begin{equation*}
\begin{split}
\mathfrak{L}^u_{\theta,\eta}(\sigma):=-i[H_0+u H_1,\sigma]+\textstyle\sum^m_{k=1}&\frac{\theta_k}{2}\Big(2(1-\eta_k)L_k \sigma L_k^*-(L^*_kL_k+\eta_k L_k^2)\sigma\\
&-\sigma(L^*_kL_k+\eta_k {L_k^*}^2)+\eta_k\mathrm{Tr}\big((L_k+L^*_k)^2\sigma\big)\sigma  \Big).
\end{split}
\end{equation*}
By the support theorem, the set $\mathcal{S}(\mathcal{H})$ is invariant for~\eqref{Eq:ODE-P} and~\eqref{Eq:ODE-F}.

\begin{proposition}\label{Prop:Recurrence}
    Suppose that $\eta_k<1$ for all $k\in[n]$ and the assumptions \emph{\textbf{AR'}}, \emph{\textbf{A1.2}}, \emph{\textbf{A1.3}}, \emph{\textbf{A2}}, \emph{\textbf{H1}}, \emph{\textbf{H2}} and \emph{\textbf{C1}} are satisfied. For all initial condition $(\sigma_0,\hat{\sigma}_0)\in\mathcal{S}(\mathcal{H})\times \mathrm{int}\{\mathcal{S}(\mathcal{H})\}$ and for any $\zeta>0$, the trajectories of the coupled system~\eqref{Eq:SME-Pf}--\eqref{Eq:SME-F} can enter $B_{\zeta}(\mathcal{H}_S)\times B_{\zeta}(\mathcal{H}_S)$ in finite time $\overline{\mathbb{P}}^{\sigma}$-almost surely for almost all values of $\alpha$ and $\gamma$.
\end{proposition}
\proof
The proof consists of four steps:
\begin{enumerate}
    \item First, we show that, for all $\hat{\sigma}_0>0$, there exists $v\in \mathcal{V}$ such that $u(\hat{\sigma}_v(t_1))\neq 0$ for some finite $t_1 \geq 0$.
    \item Next, we show that, if $\Tr(\Pi_0\sigma_v(0))=0$, there exists a finite $t_2>0$ such that $\Tr(\Pi_0\sigma_v(t_2))>0$.
    \item Then, we show that, there exists a finite $t_3>0$ and $v\in\mathcal{V}$ such that $(\sigma_v(t_3),\hat{\sigma}_v(t_3))\in B_{\zeta}(\mathcal{H}_S)\times B_{\zeta}(\mathcal{H}_S)$.
    \item Finally, we show that $(\sigma_t,\hat{\sigma}_t)$ can enter $B_{\zeta}(\mathcal{H}_S)\times B_{\zeta}(\mathcal{H}_S)$ in finite time almost surely.
\end{enumerate}

\textit{Step 1.} Suppose $u(\hat{\sigma}_v(t))= 0$ for all $v\in \mathcal{V}$ and $t\geq 0$. Then, for all $j\in [d]$, we have
\begin{align*}
\Tr(\Pi_j \dot{\hat{\sigma}}_v(t))=\textstyle 2\Tr(\Pi_j \hat{\sigma}_v(t))\sum^n_{k=1}\hat{\eta}_k\hat{\theta}_k\big(\sum^d_{i=0}\mathbf{Re}\{l_{k,i}\}^2\Tr(\Pi_i \hat{\sigma}_v(t))-\mathbf{Re}\{l_{k,j}\}^2  \big)\\
\textstyle+4\sum^n_{k=1}\sqrt{\hat{\eta}_k \hat{\theta}_k}\mathbf{P}_{k,j}(\hat{\sigma}_v(t))\Tr(\Pi_j\hat{\sigma}_v(t)) V_k(t),
\end{align*}
where $\mathbf{P}_{k,j}(\sigma)=\mathbf{Re}\{l_{k,j}\}-\sum^d_{i=0}\mathbf{Re}\{l_{k,i}\}\Tr(\Pi_i \sigma)$.
First, we consider the case $\hat{\sigma}\in \bigcap_{k\in[n]}\mathbf{S}_{k,j}\setminus B_{\varepsilon}(\mathcal{H}^j_{R})$ with $\varepsilon>0$ sufficiently small, where 
$
\mathbf{S}_{k,j}:=\{\hat{\sigma}\in\mathcal{S}(\mathcal{H})|\,\mathbf{P}_{k,j}(\hat{\sigma})=0\}.
$
In this case, we have 
\begin{align*}
    &\textstyle\sum^d_{i=0}\mathbf{Re}\{l_{k,i}\}^2\Tr(\Pi_i \hat{\sigma})-\mathbf{Re}\{l_{k,j}\}^2\\
    &=\textstyle\sum^d_{i=0}\Tr(\Pi_i \hat{\sigma})\sum^d_{i=0}\mathbf{Re}\{l_{k,i}\}^2\Tr(\Pi_i \hat{\sigma})-\big(\sum^d_{i=0}\mathbf{Re}\{l_{k,i}\}\Tr(\Pi_i \hat{\sigma})\big)^2\geq 0,
\end{align*}
where we used the fact $\sum^d_{i=0}\Tr(\Pi_i \hat{\sigma})=1$ and the Cauchy-Schwarz inequality. The last equality holds if and only if there exists $i\in [d]$ such that $\Tr(\Pi_i \hat{\sigma})=1$.
Then, we consider the second case $\hat{\sigma}\in \mathrm{int}\{\mathcal{S}(\mathcal{H})\}\setminus  \bigcap_{k\in[n]}\mathbf{S}_{k,j}$. In this case, 
$\mathbf{P}_{k,j}(\hat{\sigma})\Tr(\Pi_j\hat{\sigma})\neq 0$. Thus, by applying the similar arguments as in~\cite[Lemma 6.1]{liang2019exponential}, we can always construct a set of $v\in\mathcal{V}$ ensure $\hat{\sigma}_v(t)\in B_{\varepsilon}(\mathcal{H}^j_{R})$ for some finite $t>0$. Moreover, $u(\hat{\sigma})\neq 0$ for all $\hat{\sigma}\in B_{\varepsilon}(\mathcal{H}^j_{R})$ due to \textbf{H1}, which leads to the contradiction. 

\textit{Step 2.}
Suppose that $u(\hat{\sigma}_v(t))\neq 0$ and $\Tr(\Pi_0 \sigma_v(t))=0$ for all $t\in [t_1, t_1+\delta]$ with $\delta >0$ sufficiently small. Based on the simple linear algebra arguments, we deduce $\Pi_0 \sigma_v(t) \Pi_0 =0$ and $\sigma_v(t) \Pi_0 =0$. It implies that, for all $t\in[t_1,t_1+\delta]$
\begin{equation*}
    \Pi_0 \dot{\sigma}_v(t) \Pi_0=\textstyle\sum^n_{k=1}\theta_k(1-\eta_k) \Pi_0 L_k\sigma_v(t)L_k^* \Pi_0+\gamma \textstyle\sum^m_{j=1} \Pi_0 C_j\sigma_v(t)C_j^* \Pi_0=0.
\end{equation*}
Thus, we have $\sigma_v(t)L_k^* \Pi_0=0$ and $\sigma_v(t)C_j^* \Pi_0=0$ for all $t\in[t_1,t_1+\delta]$. Moreover, due to \textbf{AR'}, it is straightforward to show  $[\sum^m_{j=1}C^*_jC_j,\Pi_0]=0$ and $[\tilde{H},\Pi_0]=0$. Then, we obtain
\begin{equation*}
   \dot{\sigma}_v(t) \Pi_0=iu(\hat{\sigma}_v(t))\sigma_v(t) H_1 \Pi_0=0, \quad \forall t\in [t_1, t_1+\delta],
\end{equation*}
which implies $\sigma_v(t)H_1\Pi_0=0$ since $u(\hat{\sigma}_v(t))\neq 0$. Then, by applying the above arguments recursively, we have
$$
\dot{\sigma}_v(t)H_1 \Pi_0=0, \quad \dots, \dot{\sigma}_v(t)H_1 \Pi_0=0, \quad \forall t\in [t_1, t_1+\delta],
$$ 
where $l\in\mathbb{Z}$ is defined in \textbf{A2}. It implies that 
\begin{align*}
    \sigma_v(t)\Pi_0=\sigma_v(t)H_1\Pi_0=\sigma_v(t)L^*_1H_1\Pi_0=\dots=&\sigma_v(t)L^*_nH_1\Pi_0=\dots\\
    &\dots = \sigma_v(t)L^*_nH^l_1\Pi_0=0.
\end{align*}
Due to \textbf{A2}, we deduce that $\mathrm{rank}\big(\sigma_v(t)\big)\leq 1$ which leads to a contradiction, since by Lemma~\ref{Lemma:Mixed}, Lemma~\ref{Lemma:PosDef invariant} and the support theorem~\cite{stroock1972support}, $\mathrm{rank}\big(\sigma_v(t)\big)>1$ for all $t>0$.

\textit{Step 3.} From~\eqref{Eq:ODE-P}--\eqref{Eq:ODE-F}, we have, for all $t\geq t_2$, 
\begin{align*}
    \Tr(\Pi_0\dot{\sigma}_v(t)) &= \Tr\big(\Pi_0\big(\mathfrak{L}^u_{\theta,\eta}(\sigma_v(t))+\mathfrak{F}_{\alpha,\gamma}(\sigma_v(t))\big)\big)\\
    &~~~~~~~+2\textstyle\sum^n_{k=1}\sqrt{\eta_k \theta_k}\mathbf{P}_{k,0}(\sigma_v(t)) \Tr(\Pi_0\sigma_v(t)) V_k(t),\\
    \Tr(\Pi_0\dot{\hat{\sigma}}_v(t)) &= \Tr\big(\Pi_0\mathfrak{L}^u_{\hat{\theta},\hat{\eta}}(\hat{\sigma}_v(t))\big)\\
    &~~~~~~~+2\textstyle\sum^n_{k=1}\sqrt{\hat{\eta}_k \hat{\theta}_k}\mathbf{P}_{k,0}(\hat{\sigma}_v(t)) \Tr(\Pi_0\hat{\sigma}_v(t)) V_k(t),
\end{align*}
where $\Tr(\Pi_0\sigma_v(t_2))>0$. Since $\mathcal{S}(\mathcal{H})\times \mathcal{S}(\mathcal{H})$ is compact, the first two terms of the right-hand side of above two equations are bounded from above in this domain. 

Due to \textbf{A1.2} and \textbf{A1.3}, we can deduce $\bigcap^n_{k=1}\mathbf{S}_{k,0}=\mathcal{I}(\mathcal{H}_S)$. Thus, for any $\zeta>0$, there exist at least one $k\in[n]$ and $\delta>0$ such that $|\mathbf{P}_{k,0}(\sigma)|\geq \delta$ for all $\sigma\in\mathcal{S}(\mathcal{H})\setminus B_{\zeta}(\mathcal{H}_S)$.
Then by choosing $V_k(t) = K/ \min \{\mathbf{P}_{k,0}(\sigma_v(t)),\mathbf{P}_{k,0}(\hat{\sigma}_v(t))\}$, with $K > 0$ sufficiently large, we can guarantee that $(\sigma_v(t)),\hat{\sigma}_v(t))$ enter $B_{\zeta}(\mathcal{H}_S)\times B_{\zeta}(\mathcal{H}_S)$ in finite time.

\textit{Step 4.} Due to the compactness of $\mathcal{S}(\mathcal{H})\times \mathcal{S}(\mathcal{H})$ and the Feller continuity of the trajectories $(\sigma_t,\hat{\sigma}_t)$, together with Lemma~\ref{Lem:Instability}, we can conclude the proof by applying the similar arguments as in \cite[Lemma 4.10]{liang2020robustness}.
\hfill$\square$

\bibliographystyle{siamplain}
\bibliography{references}
\end{document}